\documentclass[sigconf]{acmart}

\usepackage{enumitem}
\usepackage{lipsum}
\usepackage{subcaption}
\usepackage{makecell}
\usepackage[usestackEOL]{stackengine}
\usepackage{multirow}
\usepackage{float}

\newcommand{\sg}[1]{{\color{magenta} [SG: #1]}}

\newcommand*{\metaclass}[1]{\emph{#1}} 
\newcommand*{\attr}[1]{\emph{#1}}
\newcommand*{\val}[1]{\emph{#1}}

\AtBeginDocument{%
  \providecommand\BibTeX{{%
    \normalfont B\kern-0.5em{\scshape i\kern-0.25em b}\kern-0.8em\TeX}}}

\copyrightyear{2024}
\acmYear{2024}
\setcopyright{acmlicensed}\acmConference[MODELS '24]{ACM/IEEE 27th International Conference on Model Driven Engineering Languages and Systems}{September 22--27, 2024}{Linz, Austria}
\acmBooktitle{ACM/IEEE 27th International Conference on Model Driven Engineering Languages and Systems (MODELS '24), September 22--27, 2024, Linz, Austria}
\acmDOI{10.1145/3640310.3674102}
\acmISBN{979-8-4007-0504-5/24/09}

\begin{document}

\title[Tree-Based versus Hybrid Graphical-Textual Model Editors: An Empirical Study]
% {An Empirical Study of Hybrid Graphical-Textual\\ Model Editors against Tree-Based Model Editors}
% {An Empirical Study of Testing Specification Model Editors: Tree-Based versus Hybrid Graphical-Textual}
% {An Empirical Study of Testing Specifications: Tree-Based Model Editors versus Hybrid Graphical-Textual Model Editors}
% {An Empirical Study of Testing Specifications: 
% \\Tree-Based versus Hybrid Graphical-Textual Model Editors}
{Tree-Based versus Hybrid Graphical-Textual Model Editors:
\\An Empirical Study of Testing Specifications}
%% NEW TITLE NEEDED


% \author{Ionut Predoaia}
% \affiliation{%
%   \institution{University of York}
%   \city{York}
%   \country{United Kingdom}
% }
% \email{ionut.predoaia@york.ac.uk}

% \author{James Harbin}
% \affiliation{%
%   \institution{University of York}
%   \city{York}
%   \country{United Kingdom}
% }
% \email{james.harbin@york.ac.uk}

% \author{Simos Gerasimou}
% \affiliation{%
%   \institution{University of York}
%   \city{York}
%   \country{United Kingdom}
% }
% \email{simos.gerasimou@york.ac.uk}

% \author{Christina Vasiliou}
% \affiliation{%
%   \institution{University of York}
%   \city{York}
%   \country{United Kingdom}
% }
% \email{christina.vasiliou@york.ac.uk}

% \author{Dimitris Kolovos}
% \affiliation{%
%   \institution{University of York}
%   \city{York}
%   \country{United Kingdom}
% }
% \email{dimitris.kolovos@york.ac.uk}

% \author{Antonio García-Domínguez}
% \affiliation{%
%   \institution{University of York}
%   \city{York}
%   \country{United Kingdom}
% }
% \email{a.garcia-dominguez@york.ac.uk}
% \newcommand{\authorsN}[1]{#1}
\author{Ionut Predoaia, James Harbin, Simos Gerasimou} 
\author{Christina Vasiliou, Dimitris Kolovos, Antonio García-Domínguez}
% \authorsN{Ionut Predoaia, James Harbin, Simos Gerasimou, Christina Vasiliou, Dimitris Kolovos, Antonio García-Domínguez}
\affiliation{%
  \institution{University of York, United Kingdom}
  % \city{York}
  \country{}
}
\email{ionut.predoaia, james.harbin, simos.gerasimou, christina.vasiliou, dimitris.kolovos, a.garcia-dominguez@york.ac.uk}

\renewcommand{\shortauthors}{Predoaia et al.}

\newcommand{\squishlist}{
 \begin{list}{$ullet$}
  { \setlength{\itemsep}{0pt}
     \setlength{\parsep}{3pt}
     \setlength{\topsep}{3pt}
     \setlength{\partopsep}{0pt}
     \setlength{\leftmargin}{1.5em}
     \setlength{\labelwidth}{1em}
     \setlength{\labelsep}{0.5em} } }

\newcommand{\squishlisttwo}{
 \begin{list}{$ullet$}
  { \setlength{\itemsep}{0pt}
    \setlength{\parsep}{0pt}
    \setlength{\opsep}{0pt}
    \setlength{\partopsep}{0pt}
    \setlength{\leftmargin}{2em}
    \setlength{\labelwidth}{1.5em}
    \setlength{\labelsep}{0.5em} } }

\newcommand{\squishend}{
  \end{list}  }


  %\agd{define acronyms on their first use}
\begin{abstract}
Tree-based model editors and hybrid graphical-textual model editors have advantages and limitations when editing domain models. Data is displayed hierarchically in tree-based model editors, whereas hybrid graphical-textual model editors capture high-level domain concepts graphically and low-level domain details textually. We conducted an empirical user study with 22 participants to evaluate the implicit assumption of system modellers that hybrid notations are superior, and to investigate the tradeoffs between the default EMF-based tree model editor and a Sirius/Xtext-based hybrid model editor. The results of the user study indicate that users largely prefer the hybrid editor and are more confident with hybrid notations for understanding the meaning of conditions. Furthermore, we found that the tree editor provided superior performance for analysing ordered lists of model elements, whereas activities requiring the comprehension or modelling of complex conditions were carried out faster through the hybrid editor.
\end{abstract}





%\jrh{Set CCS Concepts and Keywords}

\maketitle

\section{Introduction}
\label{SEC-INTRO}
Domain-specific languages (DSLs) can have graphical, textual, or hybrid syntaxes. Hybrid graphical-textual DSLs are languages that have a part-graphical and part-textual syntax, i.e., a hybrid graphical-textual syntax. They are particularly appropriate for cases in which one would like to use graphical representations for capturing high-level domain concepts, and textual representations for defining low-level domain details, such as behaviour and complex expressions~\cite{cooper2019engineering}. Furthermore, they are effectively used via hybrid graphical-textual model editors \cite{predoaia2023streamlining,zolotas2018towards,wei2020automatic}, i.e., editors that enable the definition of domain models via graphical and textual representations.

Throughout this paper, we use the term \textit{model editor} instead of the term \textit{workbench}; a workbench covers a broader scope, including a model editor and other components like menus, toolbars and perspectives~\cite{burnette2005eclipse}. For brevity, we use the term \textit{hybrid} instead of \textit{hybrid graphical-textual}, and the term \textit{editor} refers to a \textit{model editor}.

Choosing between a tree-based and a hybrid editor for editing domain models is not a straightforward decision, as both have advantages and drawbacks. Tree editors display domain data hierarchically and in a structured manner by exposing parent-child relationships. Diagrammatic editors contained within hybrid editors do not provide explicit structure, and they are not appropriate for displaying parent-child relationships of multiple levels. For instance, parent elements can be displayed as containers that incorporate their child elements. However, if the children have corresponding child elements, which in turn have child elements as well, displaying all these elements in a diagram can hinder comprehension, as they would effectively be displayed as a container that contains other containers, that in turn contain other containers incorporating a set of leaf child elements. Thus, in such cases, it may be preferred to use multiple diagrams, e.g., by instantiating new diagrams containing the child elements of a parent element. Consequently, the domain data is split across multiple diagrams, and users may face difficulties in being able to relate them. A common drawback of hybrid editors is the initial cost of learning the associated part-graphical and part-textual syntaxes. However, once learned, models may potentially be edited faster, considering also that typically fewer clicks are required due to the definition of textual expressions~\cite{addazi2017towards}.

Acknowledging the importance of selecting the most suitable editor to support effective and efficient modelling, in this paper we investigate the tradeoffs between tree-based and hybrid editors. System modellers often take for granted the assumption that hybrid notations are superior, however, this claim is not yet supported by empirical evidence. To evaluate this implicit assumption, we conducted an empirical user study to analyse the editor type preferred by system modellers. We recruited 22 subjects, that are either students or professionals working in industry or academia, with various levels of knowledge and expertise in computer programming, modelling, and model-driven engineering (MDE). In the user study, the participants used a DSL that enables the definition of testing specifications of robotic systems in simulation. Each participant used the DSL via both a tree editor and a hybrid editor to execute a set of tasks that were inspired by an industrial case study of the large European Commission research project SESAME (Secure and Safe Multi-Robot Systems)~\cite{sesame_website}. The user study aims to evaluate the performance and preferences of users, and provide insights on the usage of hybrid editors and tree editors.

Our empirical study revealed that users largely prefer hybrid editors and are more confident with hybrid editors for understanding the meaning of conditions. From a performance perspective, the results pinpoint that hybrid editors are superior for activities involving the comprehension or modelling of complex conditions, whereas tree editors are more efficient for tasks concerning the exploration and analysis of ordered lists of model elements.

%Furthermore, the results evidence that hybrid editors require more keystrokes and fewer clicks for modelling conditions.

\textbf{Paper structure.} Section~\ref{SEC-BACKGROUND} introduces the necessary background, Section~\ref{SEC-CASE-STUDY} describes our industrial case study, and Section~\ref{SEC-DSL} presents the fuzzing-based testing DSL. Sections~\ref{SEC-METHODOLOGY} and~\ref{SEC-RESULTS} detail the methodology and results of the user study, respectively. 
Section~\ref{SEC-THREATS} identifies threats to validity. Section~\ref{SEC-RELATED-WORK} presents related work and, finally, Section~\ref{SEC-CONCLUSIONS} concludes the paper, providing future work directions.

\section{Background}
\label{SEC-BACKGROUND}

\vspace{1mm}\noindent
\textbf{Modelling Editors.}
Hybrid editors are commonly composed of graphical diagram(s), embedded textual editor(s) with syntax-aware editing features, such as syntax highlighting, auto-completion, and refactoring, a view in which one can edit the properties of a selected model element, a view with various symbols that can be dragged and dropped on top of diagrams to create various types of model elements, and a view that consistently reports errors from the model \cite{cooper2019engineering}. Textual expressions are able to reference model elements that are defined graphically in diagrams, and consistency between the two is automatically enforced as the model evolves. 

Hybrid DSLs and their supporting editors can be developed using Graphite \cite{predoaia2023streamlining, predoaia2023towards}, a tool that facilitates the systematic engineering of hybrid editors via automatic code generation. 
Graphite (\url{https://github.com/epsilonlabs/graphite}) uses Sirius \cite{sirius_website} for the specification of the graphical part of the syntax, and Xtext~\cite{xtext_website} for defining the textual part of the syntax. Alternatively, projectional editors, such as JetBrains MPS~\cite{jetbrains_website}, can be employed for developing such hybrid editors. The research works from~\cite{scheidgen2008textual, paper_obeo_typefox, cooper2018msc, solution_altran}, which are based on hand-written code, introduce techniques for engineering such kind of hybrid editors. 
However, they do not address consistency enforcement between textual expressions and graphical model elements. Furthermore, except~\cite{cooper2018msc}, no techniques have been proposed for uniformly reporting errors. Other streams of work involve using blended modelling~\cite{ciccozzi2019blended} and Capella~\cite{capella_extension}; however,  they do not address hybrid DSLs, as justified in~\cite{predoaia2023streamlining}.

\vspace{1.2mm}\noindent
\textbf{Testing Specification. }
Multi-robot systems (MRS) are distributed and interconnected robotic
teams capable of collectively performing tasks that are beyond the
competency of a single
robot~\cite{parker2012reliability}. Simulation-based testing (SBT) enables
investigating the capacity of an MRS to operate dependably using a
virtual environment that closely resembles the real deployment
environment. Modern robotic frameworks enable realistic simulation of
robots, with strong conformance to physical hardware, making the software
deployment to real robots easier~\cite{timperley-icst18}. 
Compared to field testing,
simulation-based testing supports searching large design spaces of
components and system configurations, using limited physical hardware
resources, thus reducing significantly the overheads of detecting
potential violations of safety
requirements~\cite{harbin-models-2021,gerasimou2019towards}.

Fuzz testing is an automated software testing technique that involves providing unexpected, invalid or random data as input to a computer program, aiming to find exceptions, crashes and potential memory leaks \cite{godefroid2020fuzzing}.
Fuzz testing is considered among the most efficient approaches for finding security vulnerabilities~\cite{liang2018fuzz}. 
In this work, fuzz testing is applied as a simulation-based testing technique to measure the robustness of a robotic mission.
% , by injecting intentional errors or faulty messages into a robotic simulation to monitor its responses and measure its robustness.

% TO BE ADDED IN OTHER SECTION
%In this paper we consider a general architecture for specifying generic fuzz testing campaigns to raise the level of abstraction and make fuzz testing more comprehensible to robotics engineers, as explored in the SESAME project \cite{sesame-d6.6}. Our focus here is specifically upon user comprehensibility; exploring the accessibility of the fuzz testing configuration and results representation within a domain specific language (DSL). We quantify with a user study how easy it is for new users to understand the MRS system and testing campaigns represented, and to interact with the testing configuration and results contained within.

\section{Case Study}
\label{SEC-CASE-STUDY}

%\agd{testing does not guarantee anything: it'd be better to say that you're aiming to obtain evidence that supports an argument that the system is safe --- check wording with Simos}

%KUKA~\cite{kuka-deliverable}. 
%KUKA
The use case is an industrial case study in gearbox manufacturing
provided by our SESAME industrial partner, KUKA \cite{kuka_website}, a company that manufactures industrial robots and factory automation systems. This use case incorporates mobile robots to assemble gearboxes, together with human operators surrounding the assembly cell who assist in the process. 
During testing, we aim to obtain evidence to validate the safety requirement that robots will not generate hazards to human workers during operation. 
The primary safety consideration is to ensure that robots do not enter regions that could cause a collision hazard to human operators placed around the perimeter of the assembly cell frame. 
Under normal circumstances, motion trajectories are pre-planned to ensure conformity to this safety requirement. However, faults in communication, sensing, or motor interlock actuation may lead to robots malfunctioning and
entering these forbidden areas. 
The simulation scenario was supplied for testing by end users, and implemented using a proprietary simulation platform provided by our industrial partner and used as a digital twin~\cite{tts-dtwin-deliverable}.

\begin{figure}
    \centering
    \includegraphics[trim={0cm 5cm 0cm 6.5cm},clip,width=0.86\linewidth]{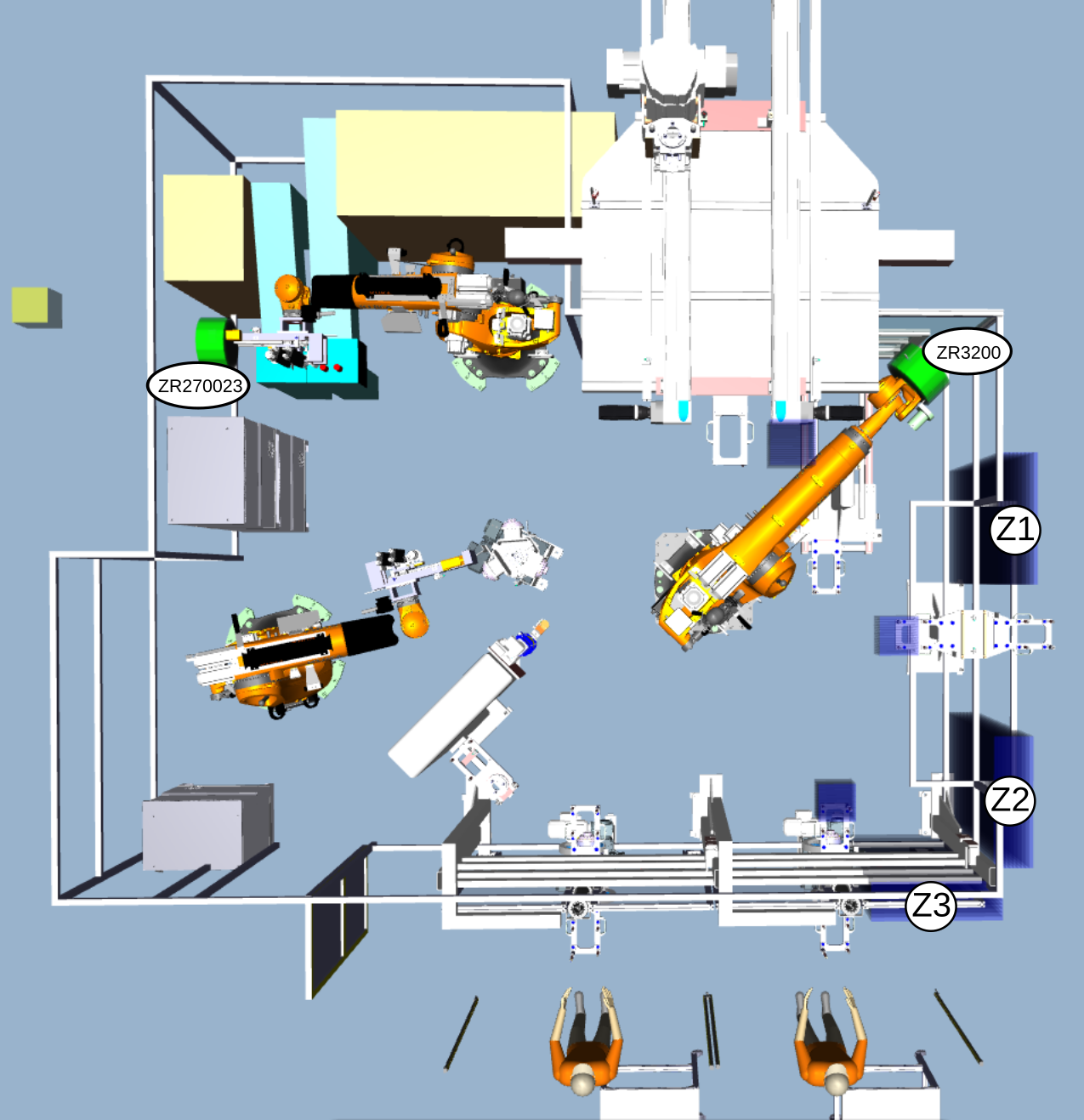}
    \vspace{-3mm}
    \caption{Case study screenshot from MRS simulation}
    \vspace{-4mm}
    \label{fig:kuka-case-study}
\end{figure}

% Individual figures version
\begin{figure}[!tb]
\centering
\includegraphics[width=\linewidth]{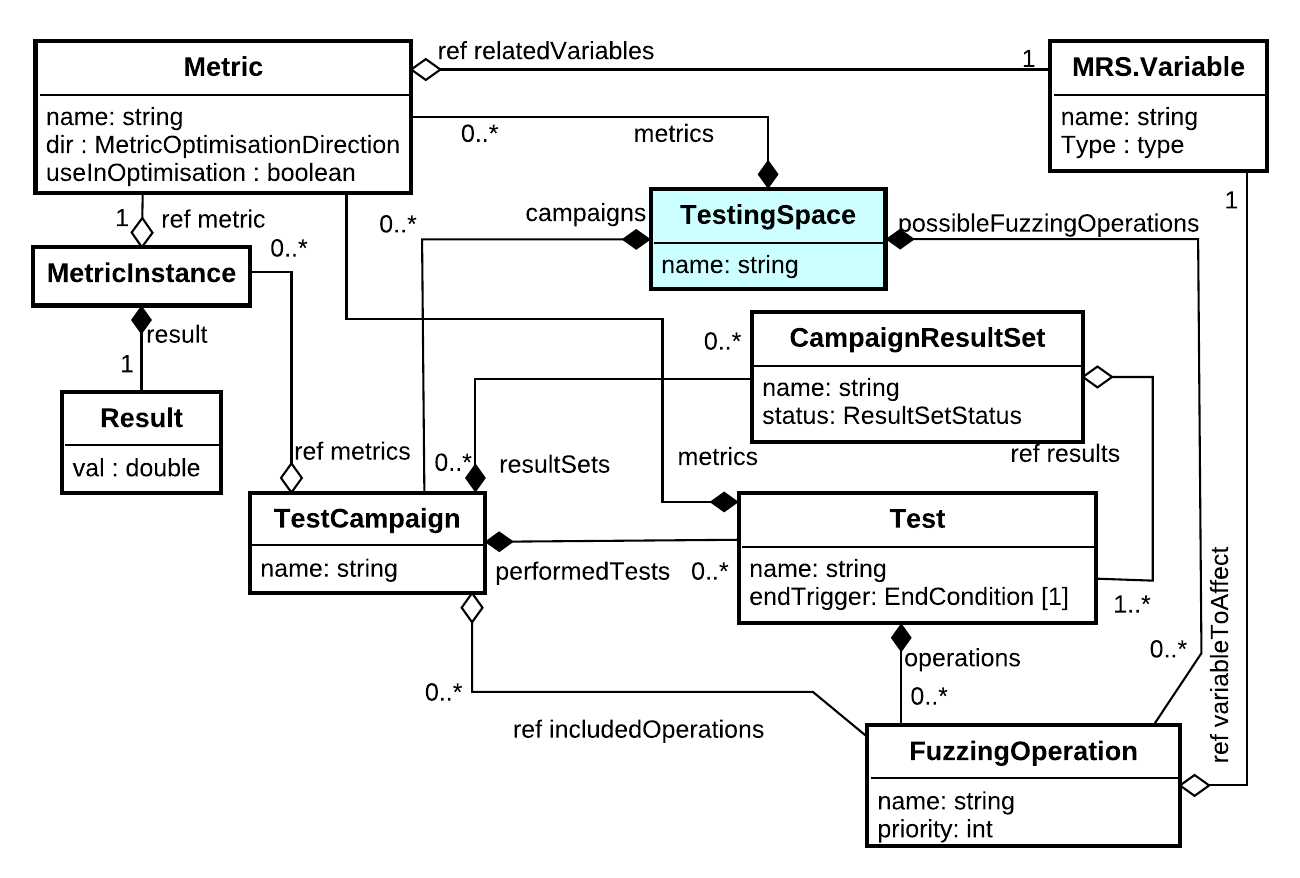}
\vspace{-8mm}
\caption{Fuzzing-based testing DSL metamodel (excerpt)}
\label{FIG-TESTSPACE-DSL}
\vspace{-3mm}
\end{figure}

\begin{figure}[!tb]
\centering
\includegraphics[width=\linewidth]{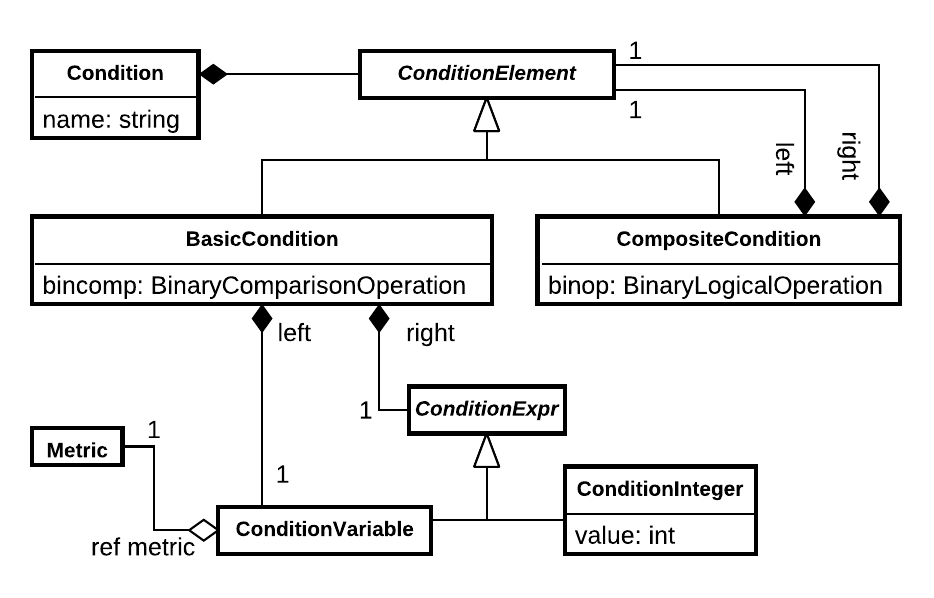}
\vspace{-9mm}
\caption{Meta-classes of condition-based fuzzing (excerpt)}
\label{FIG-CONDITIONS-DSL}
\vspace{-3mm}
\end{figure}

% JRH: subfig version of metamodel figures
\if 0
\begin{figure*}[!t]
\centering
\subfloat[][Metamodel Excerpt of the Fuzzing-Based Testing DSL]{\includegraphics[width=0.5\linewidth]{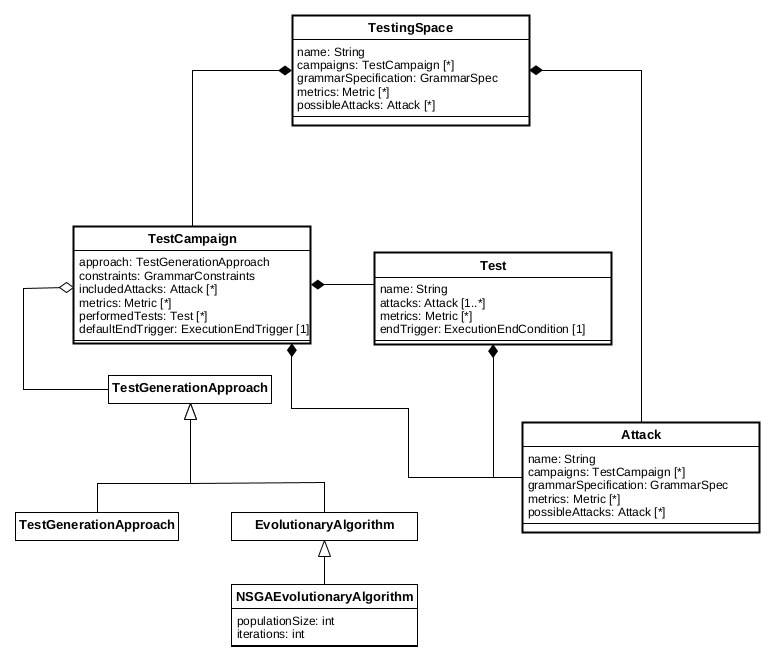}
    \vspace{-4mm}
      \label{FIG-TESTSPACE-DSL}}
\subfloat[][Meta-classes of condition-based fuzzing (excerpt)]{\includegraphics[width=0.5\linewidth]{figures/dsl-conditions-uml}
    \vspace{-4mm}
    \label{FIG-CONDITIONS-DSL}} 
    \vspace{-5mm}
\end{figure*}
\fi
  %\caption{Metamodel Excerpt of the Fuzzing-Based Testing DSL}
  %\agd{if you show an association, name it and hide the field, and add %cardinalities to associations}\agd{abstract classes should use italics}%%\agd{What are GrammarSpec and GrammarConstraints, and do we need to show it for our discussion?}\agd{How are Metrics defined?}\agd{What is the difference between ExecutionEndTrigger and ExecutionEndCondition?}\agd{Why are the results of a CampaignResultSet just references to tests? Where are their results (e.g. pass/fail)?}

Figure~\ref{fig:kuka-case-study} shows the layout of the assembly cell. 
The primary safety requirement is that the robot grippers are forbidden from entering static safety zones surrounding the cell in order to prevent hazards to human workers. 
The safety zones are illustrated as blue cuboid regions (Z1, Z2, Z3) on the right-hand side of the external cell frame in Figure~\ref{fig:kuka-case-study}. 
If the safety zone surrounding the gripper of the right-hand robot (green cylinder labelled ZR3200) collides with these regions, it will trigger violations of the static safety requirement. 
An output metric is used to quantify violations of the principal safety requirement, by tracking the occurrence of collisions of the gripper safety zone with any of Z1, Z2 or Z3. 
Following a safety analysis, we have identified the communication from the controller to the robot arm joints as vulnerable to message disruptions. 
Therefore, the primary safety requirement will be tested by applying fuzzing, altering the target motion value for robot arm joints with targeted uniform random jitter around the intended point.

We employ a DSL to raise the level of abstraction and make testing specification %and fuzz testing 
more comprehensible to robotics engineers. 
Modelling testing specifications via a DSL tailored to this specific domain is highly desirable, enabling non-expert users to easily test robotics systems~\cite{casalaro2021model,bruyninckx2013brics,harbin2023model,wood2020supporting}. 
Participants of the user study will apply the DSL in the context of this case study, via a tree editor and a hybrid editor, to investigate the following research questions:

\begin{enumerate}[font={\bfseries},label={RQ\arabic*}]
\setcounter{enumi}{0}
\item \hypertarget{link_rq1}{\textbf{(Performance)}} 
How do users perform when executing certain tasks with a hybrid editor compared to a tree editor?
\end{enumerate}

\begin{enumerate}[font={\bfseries},label={RQ\arabic*}]
\setcounter{enumi}{1}
\item \hypertarget{link_rq2}{(\textbf{Confidence})} 
How confident are users regarding the correctness and completeness of their solutions to given tasks?
\end{enumerate}

\begin{enumerate}[font={\bfseries},label={RQ\arabic*}]
\setcounter{enumi}{2}
\item \hypertarget{link_rq3}{(\textbf{Preference})} 
Do users prefer hybrid editors or tree editors?
\end{enumerate}

%TO BE REVIEWED: MAY HAVE TO MOVE THIS INTO A FOLLOWING SECTION
%A part of the fuzzing model instantiated for the KUKA industrial use case is presented in Figure~\ref{FIG-DSL-TTS-EXAMPLE}. The operations $attackJointR3200ProductMove\_Link1$ and $attackJointR3200ProductMove\_Link0$ of the model illustrate this fuzzing offset value change, using the \metaclass{RandomValueFromSet} fuzzing operation to apply relative changes to the joint value time series.
%\begin{figure}[t]
%	\centering
%        \includegraphics[width=\linewidth]{figures/tts-model-example-partial}
%	\caption{A fragment of the DSL example for the TTS use case
%          showing selected metrics and fuzzing operations}
%	\label{FIG-DSL-TTS-EXAMPLE}
%\end{figure}

\section{Fuzzing-Based Testing DSL}
\label{SEC-DSL}
This section presents the fuzzing-based testing DSL developed in the context of the SESAME project, and applied to the case study from Section~\ref{SEC-CASE-STUDY}. 
Through the DSL, users can define testing specifications for an MRS simulation, which controls how fuzzing is applied during testing.  
Next, we present the abstract syntax and the concrete syntaxes of the DSL and the testing process.

\subsection{Abstract Syntax}
\label{SEC-DSL-ABSTRACT}
The abstract syntax (i.e., the metamodel of the DSL) has been defined with the Eclipse Modelling Framework (EMF) through the Ecore metamodelling language. 
Figure~\ref{FIG-TESTSPACE-DSL} shows an excerpt from the metamodel of the DSL as a class diagram. 
The core meta-classes are \metaclass{TestingSpace}, \metaclass{TestCampaign}
and \metaclass{Test}. A \metaclass{TestingSpace} (highlighted)
represents the root element of the domain and comprises permissible fuzzing operations (\attr{possibleFuzzingOperations}) that bind
the potential fuzzing space. The testing space also includes
performance \attr{metrics}, which are used to quantify the robotic
system performance for safety violations.
% JRH: removed as this is mentioned in "Sec. 4.3: simulation-based testing"
%, and allow multi-objective optimisation of test performance
%(optimising for most impactful violations).
A \metaclass{TestCampaign} specifies and constrains an individual experiment, referencing from the \metaclass{TestingSpace} a subset of 
\attr{metrics} to use in campaign evaluation, together with a subset of chosen fuzzing operations in \attr{includedOperations}. 
The \attr{performedTests} attribute in \metaclass{TestCampaign} is automatically populated during the experiment with dynamically generated and executed \metaclass{Test}s.

%The \metaclass{TestGenerationApproach}
%allows specifying the parameters for an experiment by selecting one of
%its subclasses. For example, \metaclass{NSGAEvolutionaryAlgorithm}
%allows an evolutionary experiment with the NSGA-II
%algorithm~\cite{nsga2}. 
The \metaclass{Test} class represents a test configuration, corresponding to a specific selection of fuzzing \attr{operations}.
During \metaclass{Test} execution, instances of
\metaclass{MetricInstance} are recorded, to quantify the values of particular \metaclass{Metric}s under the applied fuzzing.

The \metaclass{FuzzingOperation} class is a specific fuzzing operation, i.e., an entity that represents a specific strategy for making runtime modifications to the
MRS. The property \attr{variableToAffect} selects the MRS variable to which fuzzing operations are applied. 
These variables will be automatically manipulated by the testing platform when fuzzing is active. 
% Not shown here; 
Subclasses of \metaclass{FuzzingOperation} represent distinct fuzzing semantics; e.g., \metaclass{RandomValueFromSetOperation}
enables randomised changes to structured parameters within a message.

\if 0
\begin{figure}[t]
\centering
\includegraphics[width=0.9\linewidth]{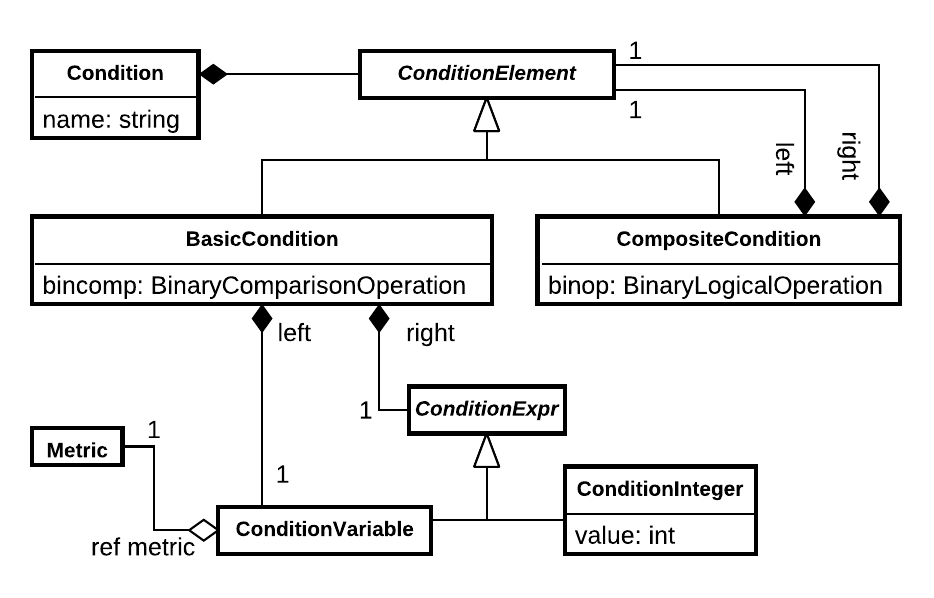}
\vspace{-4mm}
\caption{Meta-classes of condition-based fuzzing (excerpt)
\sg{see comments for polishing}
}
%\agd{we need to be consistent with capitalisation: choose one approach and stick with it}\agd{remove fields which are shown as associations, and add cardinalities to associations}\agd{shouldn't there be two compositions from CompositeCondition to ConditionElement?}}
\label{FIG-CONDITIONS-DSL}
\vspace{-3mm}
\end{figure}
\fi

%\agd{what is the Pareto front measured in?} \ip{ideally define what a Pareto front is, what it contains and in what it is measured?}.
A \metaclass{CampaignResultSet} has a \attr{name} and represents either final or partial intermediate testing \attr{results}, as specified by the \metaclass{ResultSetStatus} enumeration. The
results are shown as references to notable tests under
\attr{results}. In an evolutionary experiment, the set of
non-dominated tests in the final campaign result set constitutes the Pareto front of optimal metrics obtained during the experiment. 
The inclusion of a \metaclass{Test} in the non-dominated result sets shows that no other tests dominate it with respect to the set of \attr{metrics} used.
% Explain the metric direction and how it is used in optimisation

Condition-based fuzzing allows users to activate and deactivate fuzzing based on custom scenario-specific events, incorporating complex logic and spatial or geometric properties. 
A condition is a logical expression built hierarchically
from comparison and logical operators, incorporating state variables from the robotic system. 
The values of condition variables change dynamically during test execution, with every variable value supplied
from a \metaclass{Metric} of the corresponding name. Activation of a condition-based fuzzing operation is controlled by its start and end conditions. 
When the start condition of a fuzzing operation holds, that fuzzing operation will be activated, and will manipulate simulation events. 
Once activated, the fuzzing operation remains active until its end condition holds.

Condition-based fuzzing experiments aim to discover these
conditions automatically, identifying high-level MRS events that lead to failures or safety violations. 
The primary role of a system tester will be to inspect, parse and understand these auto-generated conditions. 
However, during debugging and exploratory testing, users may modify existing conditions or create new ones from scratch, to verify the impact of these changes on the 
behaviour of a test.

\begin{figure}
\centering
\frame{\includegraphics[width=.935\linewidth]{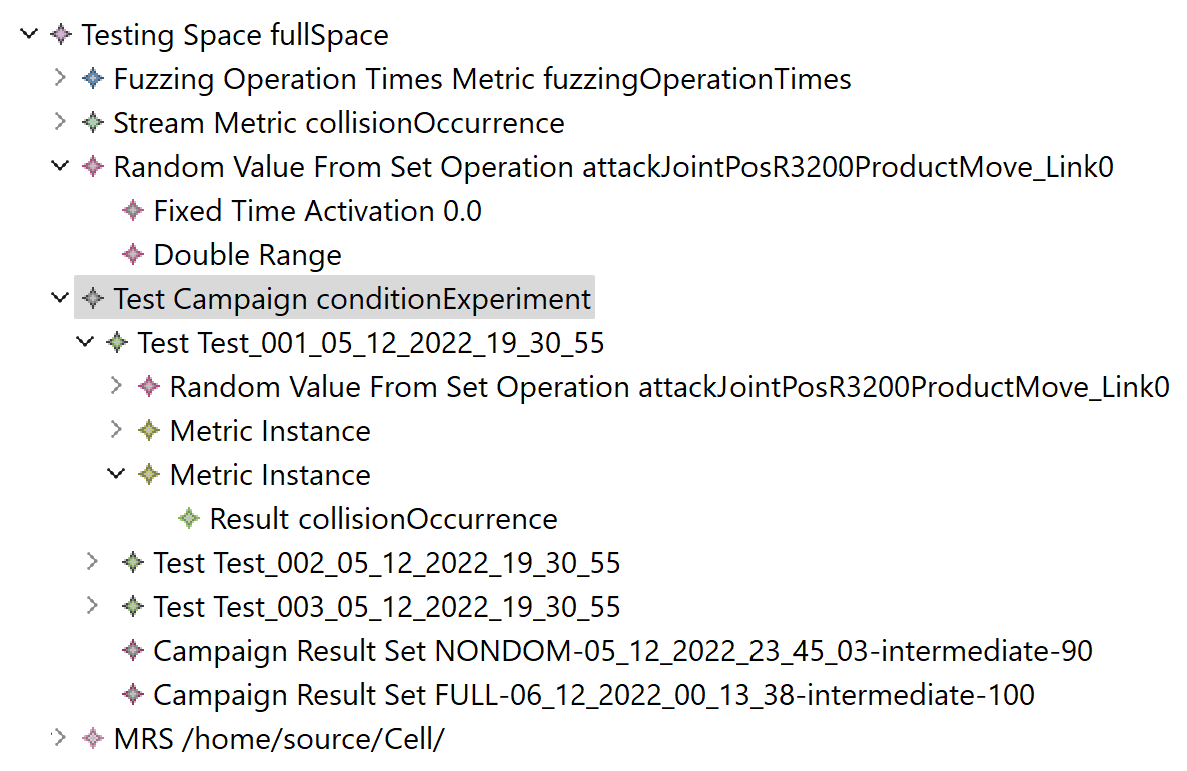}}
      \vspace{-2.9mm}
    \caption{Tree Editor --- Testing Space}
  \label{FIG-TREE-EDITOR}
    \vspace{-1.8mm}
\end{figure}

\begin{figure}[t]
\centering
\frame{\includegraphics[width=0.935\linewidth]{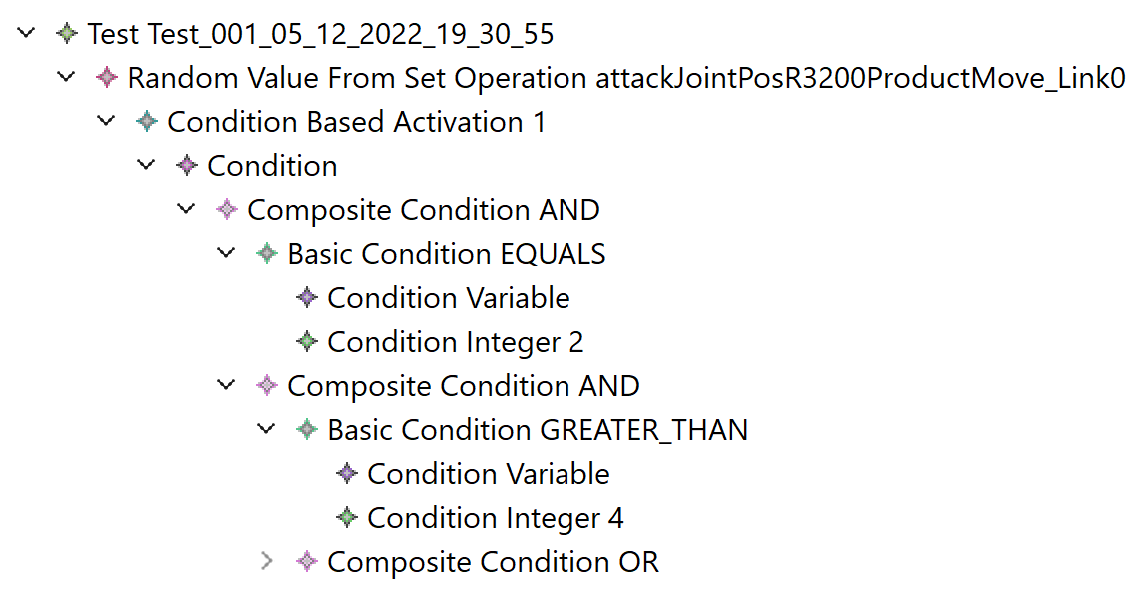}}
    \vspace{-2.9mm}
    \caption{Tree Editor --- Activation Condition}
  \label{FIG-TREE-EDITOR-CONDITIONS}
     \vspace{-1.8mm}
\end{figure}

\begin{figure}[t]
\centering
\frame{\includegraphics[width=\linewidth]{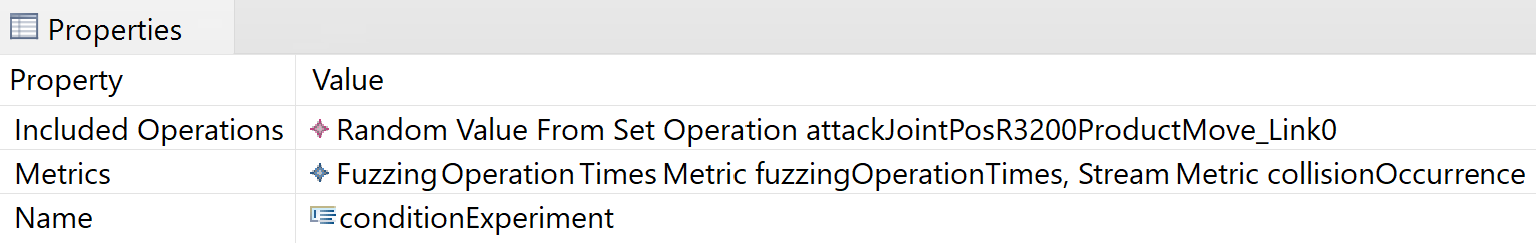}}
    \vspace{-6.8mm}
  \caption{Tree Editor --- Properties View}
  \label{FIG-TREE-EDITOR-PROPERTIES}
    \vspace{-1.8mm}
\end{figure}

\begin{figure}[t]
\centering
\frame{\includegraphics[width=\linewidth]{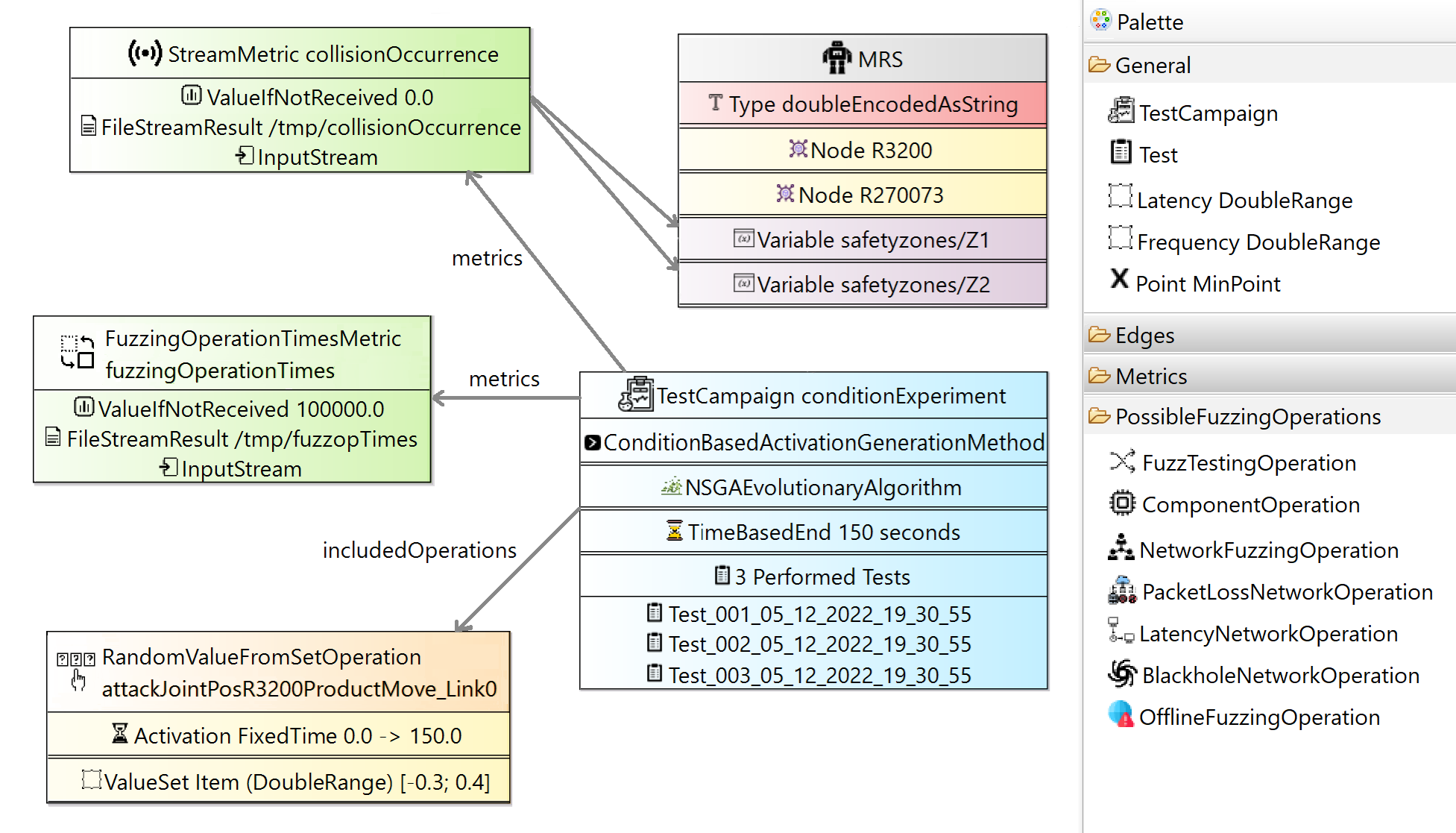}}
    \vspace{-6.8mm}
  \caption{Hybrid Editor --- Diagram of a Testing Space}
  \label{FIG-HYBRID-EDITOR-TESTINGSPACE}
  \vspace{-2.2mm}
\end{figure}

The metamodel fragment in Figure~\ref{FIG-CONDITIONS-DSL} shows the meta-classes for condition-based fuzzing. The
\metaclass{Condition} class comprises an abstract
\metaclass{ConditionElement}, which is a
\metaclass{BasicCondition} or \metaclass{CompositeCondition}. A
\metaclass{BasicCondition} represents comparisons of a
\metaclass{ConditionVariable} with a \metaclass{ConditionExpr},
which is either a \metaclass{ConditionVariable} or
\metaclass{ConditionInteger}. The
\metaclass{BinaryComparisonOperation} is either \val{LESS\_THAN},
\val{GREATER\_THAN}, or \val{EQUALS}. \metaclass{ConditionVariables}
contain a reference to an associated \attr{metric} that supplies its
value, and \metaclass{ConditionIntegers} represent primitive
integers. A \metaclass{CompositeCondition} contains two other
\metaclass{ConditionElements} in the left and right branches, together
with a \metaclass{BinaryLogicalOperation}. These logical operations
permit \val{AND} or \val{OR} operators. Since condition
expressions are dynamically generated during experiments,
\metaclass{CompositeCondition} permits complex multi-level boolean
expressions of arbitrary depth.

%TO BE REVIEWED: MAY BE MOVED TO A FOLLOWING SECTION
%In this paper, a major aspect is to investigate how different DSL syntaxes permit the comprehension of deeply nested condition expressions, allowing the user to interpret their meanings. We will also investigate how the syntaxes allow the users to manually modify these syntaxes.

\subsection{Concrete Syntaxes}

The fuzzing-based testing DSL can be used through two concrete syntaxes: a graphical tree-based syntax and a hybrid syntax. The graphical syntax is provided by the default EMF-generated tree editor, whereas the hybrid syntax was developed using Graphite~\cite{predoaia2023streamlining}.

\vspace{1.2mm}\noindent
%The testing space contains two metrics (a \metaclass{FuzzingOperationTimesMetric} and a \metaclass{StreamMetric}), a \metaclass{FuzzingOperation} of type \metaclass{RandomValueFromSetOperation} named \val{attackJointPosR3200ProductMove\_Link0}, a \metaclass{TestCampaign}, and the definition of an MRS. 
\textbf{Tree editor.} A subset of a testing space used in the case study is shown in Figure~\ref{FIG-TREE-EDITOR}. The \metaclass{TestCampaign} is selected in the editor, and its attributes (e.g., \attr{includedOperations} and \attr{metrics}) are displayed in the Properties View from Figure~\ref{FIG-TREE-EDITOR-PROPERTIES}. The test campaign contains three executed tests and two campaign result sets. Recall that a \metaclass{CampaignResultSet} represents a logical grouping of executed tests. The first test contains the \metaclass{RandomValueFromSetOperation} under \attr{includedOperations} in the \metaclass{TestCampaign}, and also two metric instances: one for \val{fuzzingOperationTimes} and one for \val{collisionOccurrence}, as specified by the \attr{metrics} attribute displayed in the Properties View from Figure~\ref{FIG-TREE-EDITOR-PROPERTIES}. Figure~\ref{FIG-TREE-EDITOR-CONDITIONS} expands the \metaclass{RandomValueFromSetOperation} named \val{attackJointPosR3200ProductMove\_Link0} and displays its child elements. This operation is activated when its start condition holds. The \metaclass{Condition} element is the root of the activation condition and it has a \metaclass{CompositeCondition}, in which the left side is a \metaclass{BasicCondition} and the right side is another \metaclass{CompositeCondition}. %Note that the \metaclass{Condition} element and its children conform to the metamodel fragment from Figure~\ref{FIG-CONDITIONS-DSL}.

% \begin{figure*}
% \centering
% \frame{\includegraphics[width=0.78\linewidth]{figures/hybrid_editor_test.png}}
%   \vspace{-2mm}
%   \caption{Hybrid Editor — Test with Condition Textual Expressions}
%   \label{FIG-HYBRID-EDITOR-TEST}
%   \vspace{-4mm}
% \end{figure*}

% \begin{figure*}
% \centering
% \frame{\includegraphics[width=0.9\linewidth]{figures/hybrid_editor_resultsets.png}}
%   \vspace{-2mm}
%   \caption{Hybrid Editor — Diagram of Campaign Result Sets}
%   \label{FIG-HYBRID-EDITOR-RESULTSETS}
%   \vspace{-1mm}
% \end{figure*}

\begin{figure*}

\hspace{2mm}
\begin{subfigure}{0.705\textwidth}
    \frame{\includegraphics[height=4.7cm]{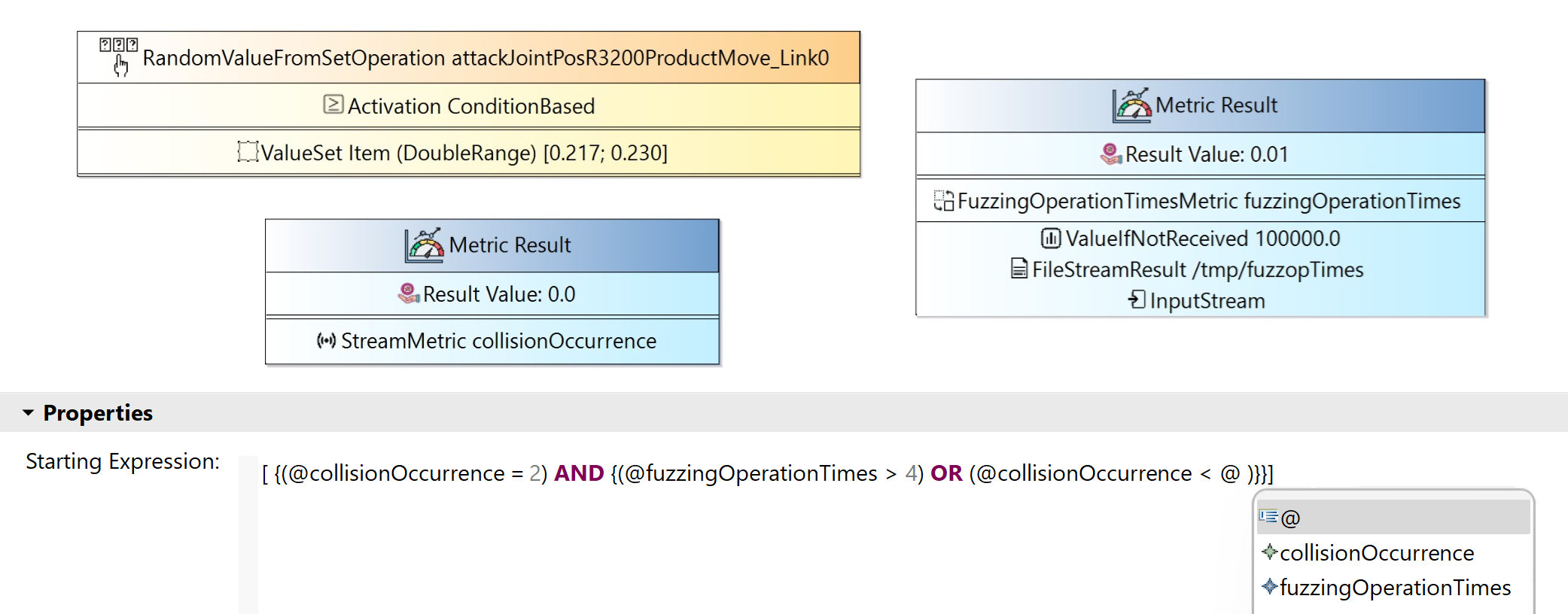}}
    
  % \caption{Hybrid Editor — Test with Condition Textual Expressions}
  % \label{FIG-HYBRID-EDITOR-TEST}
\end{subfigure}
\begin{subfigure}{0.24\textwidth}
    \frame{\includegraphics[height=4.7cm]{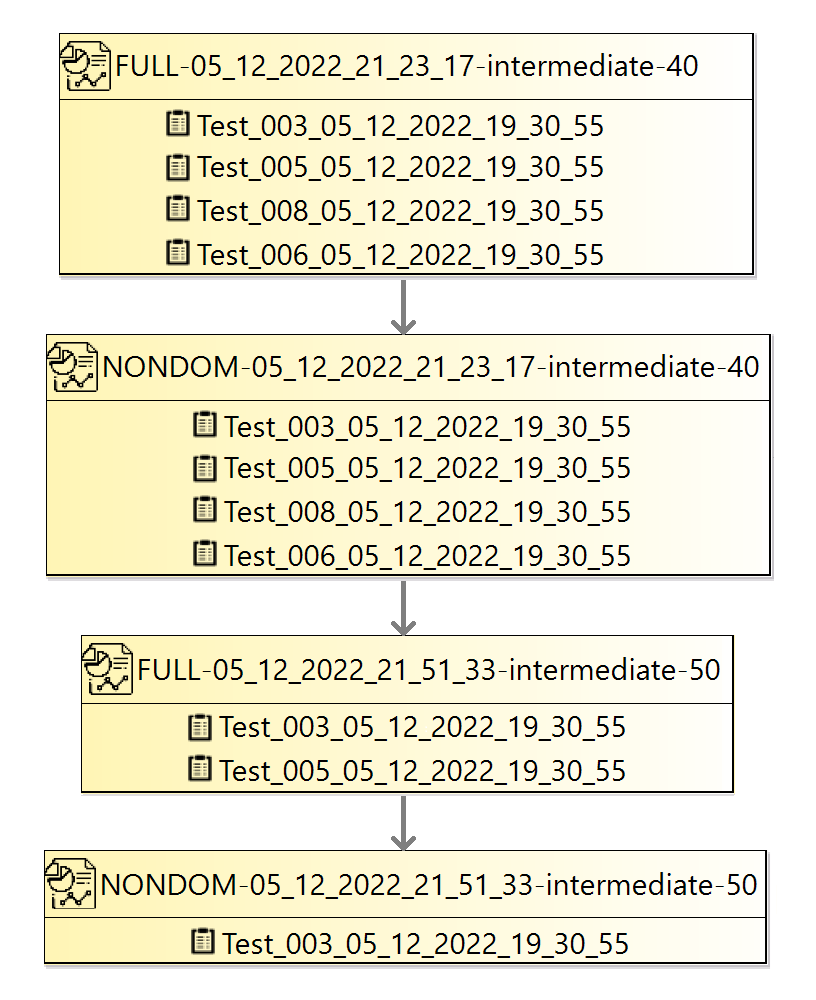}}

  % \caption{Hybrid Editor — Diagram of Campaign Result Sets}
  % \label{FIG-HYBRID-EDITOR-RESULTSETS}
\end{subfigure}
  
\vspace{-2mm}
\caption{Hybrid Editor --- a test with a condition textual expression (left diagram) and campaign result sets (right diagram)}
\label{FIG-HYBRID-EDITOR-RESULTSETS}
\vspace{-2mm}
\end{figure*}

\vspace{1.2mm}\noindent
\textbf{Hybrid editor.} The Graphite-based concrete syntax is shown in Figures~\ref{FIG-HYBRID-EDITOR-TESTINGSPACE} and %, \ref{FIG-HYBRID-EDITOR-TEST}, and~
\ref{FIG-HYBRID-EDITOR-RESULTSETS}. 
A graphically defined testing space is illustrated in the diagram in Figure~\ref{FIG-HYBRID-EDITOR-TESTINGSPACE}. The elements of the testing space are displayed as rectangles with various colours and symbols depending on each element type. Arrows highlight relationships between elements. To create new elements, users can drag and drop various symbols from the Palette on top of the diagram. The diagram displays the high-level elements of the testing space; however, users can view lower-level details by double-clicking on specific elements. 
By double-clicking on a \metaclass{Test} contained in a \metaclass{TestCampaign}, a new diagram is instantiated with the child elements of the \metaclass{Test}. For instance, Figure~\ref{FIG-HYBRID-EDITOR-RESULTSETS} (left diagram) shows a hybrid editor that contains the child elements of the first test from the test campaign. The activation condition is specified with a textual expression in an embedded textual editor, with developer assistance features such as syntax highlighting and auto-completion. Users can also double-click the \metaclass{TestCampaign} from Figure~\ref{FIG-HYBRID-EDITOR-TESTINGSPACE} to view its campaign result sets in a newly instantiated diagram. Figure~\ref{FIG-HYBRID-EDITOR-RESULTSETS} (right diagram) illustrates the campaign result sets of the test campaign, ordered from first to last via arrows. By double-clicking a \metaclass{Test} from the campaign result sets, a new diagram similar to the one from Figure~\ref{FIG-HYBRID-EDITOR-RESULTSETS} (left diagram) is instantiated, illustrating the details of the \metaclass{Test}. 
We should note that the capabilities and behaviour of the hybrid editor, e.g., the instantiation of new diagrams via double-clicking, are shown to the participants of the user study via a video tutorial.

\subsection{Simulation-Based Testing}
The SESAME tools for simulation-based testing \cite{sesame-deliverable-d6.2,sesame-deliverable-d6.6} provide a model-driven
code generation engine. Input models conforming to the DSL configure an evolutionary optimisation
loop for a given \metaclass{TestCampaign}. The loop optimises a
population of fuzzing tests with the intent of maximising safety
requirement violations. For every \metaclass{Test} generated and added
to the model, the SESAME code generation engine generates a test
runner, including an MRS logical interface enabling two-way runtime
communication. The test runner launches the simulation,
monitors and manipulates fuzzing variables to perform the operations
defined for that test. 
% \jrhchanged{When a fuzzing operation is active, all messages of the target variable will be modified}, delayed or deleted as specified by the fuzzing operation. 
Performance metrics are computed from monitored variables, and their final values are logged under
\metaclass{Test} as a \metaclass{MetricInstance}.

\begin{table}[t]
\begin{center}
\begingroup
\renewcommand*{\arraystretch}{0.7}
\caption{\label{tab:demographics}Breakdown of expertise in participants}
\vspace{-3mm}
\begin{tabular}{|>{\centering}m{1.5cm}|>{\centering}m{1.5cm}|
>{\centering}m{1.2cm}|
>{\centering}m{1.2cm}|>{\centering\arraybackslash}m{1.2cm}|} 
\hline
\vspace{1.1mm}
\textbf{Expertise} & \vspace{1.1mm} \textbf{Modelling} & \vspace{1.1mm} \textbf{DSLs} & \vspace{1.1mm} \textbf{MDE} & \vspace{1.1mm} \textbf{MRS} \\[1.3mm]
\hline
\vspace{0.7mm} None\;/\;Low & \vspace{0.75mm} 6 & \vspace{0.75mm} 7 & \vspace{0.75mm} 8 & \vspace{0.75mm} 17 \\
\hline
\vspace{0.7mm} Medium & \vspace{0.75mm} 6 & \vspace{0.75mm} 8 & \vspace{0.75mm} 4 & \vspace{0.75mm} 3 \\
\hline
\vspace{0.7mm} High & \vspace{0.75mm} 10 & \vspace{0.75mm} 7 & \vspace{0.75mm} 10 & \vspace{0.75mm} 2 \\
\hline
\end{tabular}
\endgroup
\end{center}
% \vspace{2mm}
\vspace{-4mm}
\end{table}

\section{Study Methodology}
\label{SEC-METHODOLOGY}
% To answer the questions, w
% We conducted an empirical study, recruiting individuals to test the DSL concrete syntaxes for a diverse set of scenarios and tasks.

\subsection{Recruitment}
As the study was part of a wider research project with a broad range of industry and academic partners in modelling platforms, we emailed an invitation letter to target individuals with relevant experience, with a request to indicate their availability for the 2-hour-long study. We shared the same email invite within the Computer Science department of the University of York, UK, for undergraduate and postgraduate students and academics with experience in computer programming, modelling or MDE. The invitation was also shared with various research groups in academia and professionals working in the industry, with modelling expertise.

\subsection{Participants}
A total of 22 participants participated in the study. The participants currently work either in the industry or academia: 5 participants work in the industry (3 companies across the UK, Germany and Romania), and 17 participants come from academia across 8 universities in the UK, Spain, Austria and Germany. Two of the academic participants were computer science students. Participants had a range of years of programming experience, with 10 participants with more than 10 years, 5 participants with 7--10 years, 5 participants with 4--6 years, and 2 participants with less than 1. Table \ref{tab:demographics} shows the level of knowledge and expertise of participants.

\subsection{Procedures}
The study was pilot-tested with two users who did not participate in the final study. During the pilot, we identified practical issues and revised the study accordingly, e.g., to reduce the time needed for the study and improve the wording of task descriptions. 
Once a participant agreed to join the study, the first author would contact them to arrange a time for the user study. 
A facilitator, i.e., the first author, guided and observed participants during the sessions. The faculty research committee ethically approved the study.

Each participant joined a virtual meeting with video conferencing together with the facilitator via Zoom. %\footnote{a video conferencing tool that allows users to meet online through audio and video}. 
At the beginning of the session, the facilitator described the context of the user study and shared two documents with the participant: an information sheet describing the study and a consent form. The participant read these documents and then signed the consent form if they agreed to continue with the user study session. 
Next, the facilitator shared a 20-minute video tutorial, presenting the abstract and concrete syntaxes of the fuzzing-based testing DSL and practical examples of its usage in the Eclipse Integrated Development Environment (IDE) via the hybrid and tree editors. After watching the video, the participants could ask questions on the presented information.

Each participant completed a Google Forms %\footnote{web-based software for surveys and questionnaires that collects and stores responses} 
survey, which contained the tasks %- for participants to complete - 
and follow-up questions (e.g., regarding the correctness and confidence in their answers/solutions). Throughout the user study, the facilitator clarified any uncertainties related to the task descriptions. After completing all tasks, participants were asked to report their preferences, including features they liked and disliked about the editors. Lastly, the participants completed a demographic survey, including details of their years of experience and level of expertise in the areas of Table~\ref{tab:demographics}.%  such as computer programming, modelling, MDE, DSLs and robotics software.

The participant connected remotely to the computer of the facilitator via TeamViewer. %\footnote{a tool that allows users to remotely access and control computers}. 
The computer ran a Windows Operating System (OS). On the computer, two Eclipse IDE instances were running for defining and editing testing specifications: one containing the tree editor and another containing the hybrid editor. Participants executed given tasks through both editors using the same semantic model.
To reduce memorising a task solution,
% prevent them from memorising 
participants were not told that the underlying semantic model is the same.

To reduce bias and raise the generalisability of the results, the order in which the two types of editors were used for a specific task was interchanged between participants. Half of the participants first carried out a given task with the hybrid editor, and then performed the same task once more with the tree editor. For the other half, the order was reversed, thus participants first executed the task with the tree editor. By alternating the order we can reduce its effects on performance, confidence, or preference, as a task is typically performed more efficiently and faster the second time than the first.

To meticulously analyse how users performed, all sessions were audio- and video-recorded.
The participants were encouraged to think aloud while completing the tasks, to record and later analyse any relevant comments. 
Also, a tool for visually highlighting mouse clicks was used to display a red-coloured circle around the mouse pointer when a mouse click occurred. This allows for counting only relevant mouse clicks when analysing the video recordings. For example, mouse clicks outside the editors are not relevant.

\begin{table}[t]
\begin{center}
\begingroup
\renewcommand*{\arraystretch}{0.888}
\caption{\label{tab:tasks}Tasks for Understanding (U) and Modelling (M)}
\vspace{-3mm}
\begin{tabular}{|>{\centering}m{0.5cm}|>{\raggedright\arraybackslash}m{7.2cm}|} 
\hline
\vspace{1mm} \textbf{{ID}} & \vspace{1mm} \textbf{{Description}} \\[1mm]
\hline
\vspace{0.5mm} \hypertarget{U1}{U1} & \vspace{0.5mm} In the first test, find the \metaclass{Result Value} of the \metaclass{Metric Result} associated with \metaclass{FuzzingOperationTimesMetric}.  \\
\hline
\vspace{0.5mm} \hypertarget{U2}{U2} & \vspace{0.5mm} In the first test, in the \metaclass{RandomValueFromSetOperation}, describe the activation starting condition. \\
\hline
\vspace{0.5mm} \hypertarget{U3}{U3} & \vspace{0.5mm} In the campaign result sets of the \metaclass{TestCampaign}, find the differences among the last 3 \val{NONDOM} sets. \\
\hline
\vspace{0.5mm} \hypertarget{U4}{U4} & \vspace{0.5mm} In the testing space, find the variable(s) related to the \metaclass{StreamMetric} named \val{collisionOccurrence}. \\
\hline
\vspace{0.5mm} \hypertarget{U5}{U5} & \vspace{0.5mm} In the testing space, find the fuzzing operation with the largest \metaclass{DoubleRange} interval. \\
\hline
\vspace{0.5mm} \hypertarget{M1}{M1} & \vspace{0.5mm} In the testing space, create a new \metaclass{TestCampaign} and set its name as \val{NewTestCampaign}. \\[0.25cm]
\hline
\vspace{0.5mm} \hypertarget{M2}{M2} & \vspace{0.5mm} Copy the first test from the prior existing \metaclass{TestCampaign} into the \val{NewTestCampaign}. \\
\hline
\vspace{0.5mm} \hypertarget{M3}{M3} & \vspace{0.5mm} In the \metaclass{RandomValueFromSetOperation}, modify the activation starting condition by replacing the operators. \\
\hline
\vspace{0.5mm} \hypertarget{M4}{M4} & \vspace{0.5mm} In the \metaclass{RandomValueFromSetOperation}, change the first referenced variable of the activation starting condition. \\
\hline
\vspace{0.5mm} \hypertarget{M5}{M5} & \vspace{0.5mm} Define from scratch the activation ending condition for the \metaclass{RandomValueFromSetOperation}. \\
\hline
\end{tabular}
\endgroup
\end{center}
\vspace{-4.484mm}
\end{table}

\subsection{Tasks}
Participants were given the tasks from Table~\ref{tab:tasks}, categorised into two types. 
Tasks \hyperlink{U1}{U1}--\hyperlink{U5}{U5} focus on understanding a testing space, used to evaluate which type of model editor is better for comprehension.
Tasks \hyperlink{M1}{M1}--\hyperlink{M5}{M5} focus on modelling a testing space, used for evaluating which model editor provides better performance when modelling.
The table shows a summary of the tasks; the complete task descriptions used in the study are available in \cite{reproduction_package}.
%The tasks for understanding a testing space are used to evaluate which type of model editor is better for comprehension. 
% Correspondingly, the tasks for modelling a testing space are used for evaluating which model editor provides superior performance when modelling.
All given tasks were driven by the industrial case study, aiming to reflect realistic modelling scenarios performed in practice by domain experts while covering a wide range of modelling activities such as finding specific elements, finding relationships and differences between elements, changing references, and creating and copying elements. Participants operate over the semantic model presented in Figure~\ref{FIG-TREE-EDITOR}, which contains a subset of the relevant model elements from the actual model used in the case study. All tasks of the same type were first executed with one editor, and then they were executed again with the other editor, e.g., after completing all modelling tasks with the hybrid editor, they were executed again with the tree editor.

\vspace{1mm}\noindent 
\textbf{Understanding Tasks (\hyperlink{U1}{U1}\,--\,\hyperlink{U5}{U5}).}
To perform the tasks using the tree editor, participants have to carefully inspect all model elements, and accordingly expand them to view their child elements and scan their attributes displayed in the Properties View. We describe next how a user should execute the tasks via the hybrid editor. 

Task \hyperlink{U1}{U1} requires participants to locate various model elements with specific values of interest. More specifically, the task asks participants to locate the first test contained in the test campaign from Figure~\ref{FIG-HYBRID-EDITOR-TESTINGSPACE} and view its details through a newly instantiated diagram by double-clicking the test. In the instantiated diagram, that looks like that in Figure~\ref{FIG-HYBRID-EDITOR-RESULTSETS}, participants must locate the \metaclass{Metric Result} that contains a \metaclass{FuzzingOperationTimesMetric}, and then identify its \metaclass{Result Value}. For the diagram from Figure~\ref{FIG-HYBRID-EDITOR-RESULTSETS}, the \metaclass{Result Value} is 0.01.

Task \hyperlink{U2}{U2} asks participants to describe conditions defined through a tree representation and a textual representation. The task entails viewing the details of the first test as presented above, locating the \metaclass{RandomValueFromSetOperation}, finding the activation starting condition, and describing it. 
In the tree editor, the activation starting condition looks as in Figure~\ref{FIG-TREE-EDITOR-CONDITIONS}, whereas in the hybrid editor, it looks as the textual expression displayed in the embedded textual editor from Figure~\ref{FIG-HYBRID-EDITOR-RESULTSETS}. When using the tree editor, the participant was asked to construct an equivalent textual expression as pseudocode. Correspondingly, when using the hybrid editor, the participant was asked to find the textual expression, describe its meaning, and categorise conditions as either \metaclass{BasicCondition} or \metaclass{CompositeCondition}, and identify the left and right branches of \metaclass{CompositeConditions}. 

Task \hyperlink{U3}{U3} asks participants to report differences or similarities among an ordered list of model elements. 
Participants should locate the \metaclass{TestCampaign} from Figure~\ref{FIG-HYBRID-EDITOR-TESTINGSPACE} and analyse its campaign result sets using a newly instantiated diagram by double-clicking the \metaclass{TestCampaign}. In the instantiated diagram, like that in Figure~\ref{FIG-HYBRID-EDITOR-RESULTSETS}, participants must find the last 3 campaign result sets whose names are prefixed with \val{NONDOM}, and then notice their differences and similarities. 
Figure~\ref{FIG-HYBRID-EDITOR-RESULTSETS} contains a subset of campaign result sets from those used in the user study, which included 15 result sets.

Task \hyperlink{U4}{U4} asks participants to find relationships (references) between elements. 
Participants should find the \val{collisionOccurrence} \metaclass{StreamMetric} from Figure~\ref{FIG-HYBRID-EDITOR-TESTINGSPACE}, and then identify the related variables. For instance, as indicated by the arrows in Figure~\ref{FIG-HYBRID-EDITOR-TESTINGSPACE}, the \metaclass{StreamMetric} is related to the variables \val{safetyzones/Z1} and \val{safetyzones/Z2}.

Task \hyperlink{U5}{U5} asks participants to find specific types of model elements and perform minor mental computations. Participants should locate the fuzzing operation with the largest \metaclass{DoubleRange} interval. Note that the largest interval is that with the highest difference between the upper bound and the lower bound. Although multiple fuzzing operations existed in the testing space of the user study, in Figure~\ref{FIG-HYBRID-EDITOR-TESTINGSPACE} there is only one fuzzing operation: the \metaclass{RandomValueFromSetOperation}, which has a \metaclass{DoubleRange} interval of [-0.3; 0.4].

\vspace{1.2mm}
\noindent 
\textbf{Modelling Tasks (\hyperlink{M1}{M1}\,--\,\hyperlink{M5}{M5}). }
In these tasks, participants edit existing and create new model elements, and change relevant property values via both editors. The modelling tasks are done after completing the understanding tasks; thus, participants have a good understanding of the testing space before engaging in these tasks.

Task \hyperlink{M1}{M1} requests participants to create a new model element and set one of its properties.  For instance, in the hybrid editor from Figure~\ref{FIG-HYBRID-EDITOR-TESTINGSPACE}, the participant would have to identify in the Palette the symbol related to \metaclass{TestCampaign}, then drag and drop it on the diagram, and lastly set its name using the Properties View.

Task \hyperlink{M2}{M2} requires participants to execute operations within the model editors, specifically, the copy and paste operations. The participant would have to either use the context menu or a keyboard shortcut to trigger the copy operation of the first test, and then execute the paste operation into the new test campaign.

Task \hyperlink{M3}{M3} asks participants to modify an existing condition, through its tree representation as in Figure~\ref{FIG-TREE-EDITOR-CONDITIONS} and its textual representation as in Figure~\ref{FIG-HYBRID-EDITOR-RESULTSETS}. More specifically, the task requires the participant to replace all \val{AND} operators with \val{OR} operators, and replace the comparison operators \val{LESS\_THAN} and \val{EQUALS} with \val{GREATER\_THAN}. 

Task \hyperlink{M4}{M4} requests participants to change a reference within a condition. 
In particular, the task involves changing the first referenced metric from the activation starting condition with another metric. For instance, the first referenced metric in the condition from Figure~\ref{FIG-HYBRID-EDITOR-RESULTSETS} is \val{collisionOccurrence}, and participants had to use \val{fuzzingOperationTimes} instead. 
This change would be realised in the tree editor by changing the reference via a selection menu, whereas in the hybrid editor, this would be done by either typing and deleting text manually, or by using the auto-completion menu that lists all metrics that can be referenced.

Task \hyperlink{M5}{M5} requires participants to define a condition from scratch that meets certain requirements. 
Participants should define the activation ending condition such that it contains one root \metaclass{CompositeCondition} containing two \metaclass{BasicConditions}. When defining conditions as in Figure~\ref{FIG-TREE-EDITOR-CONDITIONS}, multiple clicks are required to create new elements, whereas via the hybrid editor users can leverage the auto-completion menu to define the conditions textually.

\vspace{-1mm}
\subsection{Data Collection and Analysis}

We analysed the video recordings and  the surveys completed by the participants to derive the following metrics:
% Various metrics have been collected from the video recordings and from the surveys completed by the participants, as follows:

\vspace{1mm}\noindent\textbf{Duration}. The duration quantifies the amount of effort required to do a task, measuring the time-on-task, in seconds, within the editors. 
The duration does not include time spent reading task descriptions, or
% It does not include time spent reading task descriptions. 
time wasted when the remote access and control session via TeamViewer was lost. %Note that 
Participants alternated %switched back and forth 
between the Google Forms survey and the Eclipse IDE instances containing the editors, thus we were able to monitor the time spent on each.

\vspace{0.8mm}\noindent
\textbf{Correctness}. After performing all user study sessions, we examined whether participants completed the tasks correctly or made mistakes. 
% we noticed that almost all participants completed the tasks correctly and completely. Otherwise, the participants made minor mistakes. 
Thus, we marked solutions as either 1 (correct and complete) or 0 (mistakes). When a task was not solved correctly, we recorded the specific mistake(s) made.

\vspace{0.8mm}\noindent
\textbf{Clicks and Keystrokes}. Mouse clicks and keystrokes reflect modelling effort~\cite{hill2011measuring}. The number of clicks and keystrokes was recorded for the tasks related to conditions modelling, i.e., tasks \hyperlink{M3}{M3} and \hyperlink{M5}{M5}.

\vspace{0.8mm}\noindent
\textbf{Confidence}. After performing each task, participants self-reported their confidence level in their solution, ranging from 1 (very low confidence) to 5 (very high confidence).

\vspace{0.8mm}\noindent
\textbf{Errors Corrected}. We recorded the mistakes made in the context of a modelling task, which were later corrected by the user during the execution of the same task and before its conclusion. A mistake represents a performed action that is incorrect, redundant or certainly unrelated to the solution of a modelling task. %This metric was collected for task \hyperlink{M5}{M5}, where conditions are modelled from scratch. 
Note that the clicks and keystrokes and the time spent while making mistakes and correcting them were included in the analysis.

\section{Results and Discussion}
\label{SEC-RESULTS}
% We discuss the results of the user study, considering performance, confidence and preference. 
% Also, we provide aggregated insights and preferences of participants on the usage of the model editors.

\subsection{Performance (RQ1)}
We evaluate research question \hyperlink{link_rq1}{RQ1} when carrying out various tasks using the tree-based and hybrid editors by considering the metrics: duration, correctness, errors corrected and clicks and keystrokes.

% Research question RQ1 aims to evaluate the performance of 

\vspace{1.2mm}\noindent
\textbf{Duration.} 
Table~\ref{tab:durations_statistics} shows the absolute differences between the durations for completing the tasks via the tree ($T$) and hybrid ($H$) editors. The table is split into tasks for understanding (U) and modelling (M), and the five numeric columns indicate tasks \hyperlink{U1}{U1}--\hyperlink{U5}{U5} and \hyperlink{M1}{M1}--\hyperlink{M5}{M5}. 
The delta ($\Delta$) subtracts the duration when using the hybrid editor from the duration taken with the tree editor, i.e., $Delta = Duration(Tree) - Duration(Hybrid)$. For each task, the delta of the mean and median durations are displayed. A positive value signifies a higher duration with the tree editor compared to the hybrid editor; the opposite holds for a negative value.
In addition, Figure~\ref{FIG-DURATION-BOXPLOTS} shows boxplots with the duration needed for executing the understanding and modelling tasks using both editors. 

\begin{figure*}[t!]
\centering
\includegraphics[width=0.99\linewidth]{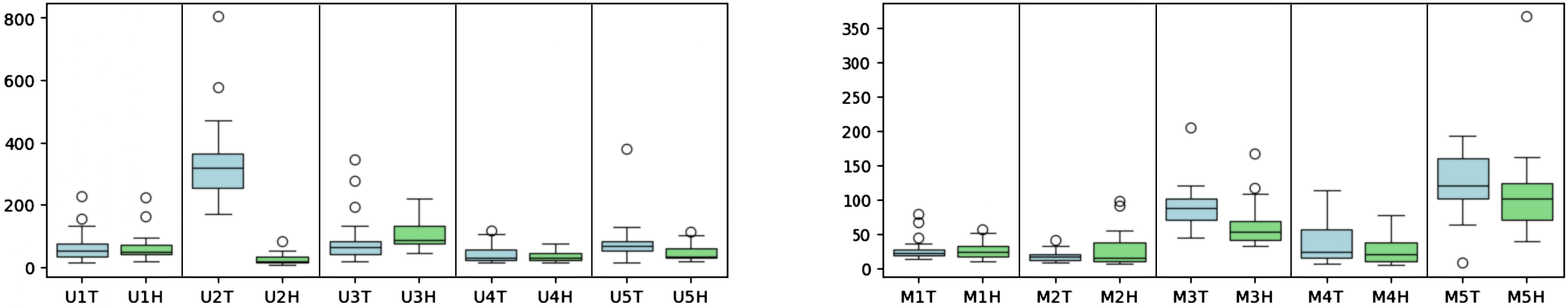}
  \vspace{-2.5mm}
  \caption{Boxplots --- Durations (seconds) for understanding (U) and modelling (M) tasks using the tree (T) and hybrid (H) editors}
  \label{FIG-DURATION-BOXPLOTS}
  \vspace{-2mm}
\end{figure*}

\begin{table}[t]
\begin{center}
\begingroup
\caption{\label{tab:durations_statistics}Deltas of Mean and Median Durations (seconds)}
\vspace{-3mm}
\renewcommand*{\arraystretch}{0.5}
\begin{tabular}{|>{\centering}m{0.7cm}|>{\centering}m{1.3cm}|
>{\centering}m{0.77cm}|
>{\centering}m{0.77cm}|
>{\centering}m{0.77cm}|
>{\centering}m{0.77cm}|>{\centering\arraybackslash}m{0.77cm}|} 
\hline
\vspace{1.2mm}
\textbf{{Type}} & \vspace{1.2mm} $\mathbf{\Delta}$ $(\boldsymbol{T}-\boldsymbol{H})$  & \vspace{1.2mm} \textbf{{\#1}} & \vspace{1.2mm} \textbf{{\#2}} & \vspace{1.2mm} \textbf{{\#3}} & \vspace{1.2mm} \textbf{{\#4}} & \vspace{1.2mm} \textbf{{\#5}} \\[1.8mm]
\hline
%\vspace{2mm} \multirow{2}{*}{\textbf{U}}
\multirow{2}{*}{\parbox{1\linewidth}{\vspace{2mm} \textbf{\hspace{2.3mm}U}}}
 & \vspace{0.8mm} Mean & \vspace{0.8mm} 1.55 & \vspace{0.8mm} 315.95 & \vspace{0.8mm} -15.32 & \vspace{0.8mm} 10 & \vspace{0.8mm} 36.45 \\[1mm]
\cline{2-7}
& \vspace{0.8mm} Median & \vspace{0.8mm} 1.50 & \vspace{0.8mm} 301 & \vspace{0.8mm} -23 & \vspace{0.8mm} 0.50 & \vspace{0.8mm} 32.50 \\[1mm]
\hline
\multirow{2}{*}{\parbox{1\linewidth}{\vspace{2mm} \textbf{\hspace{2.1mm}M}}} & \vspace{0.8mm} Mean & \vspace{0.8mm} 0.59 & \vspace{0.8mm} -9.50 & \vspace{0.8mm} 27.09 & \vspace{0.8mm} 8.94 & \vspace{0.8mm} 9.51 \\[1mm]
\cline{2-7}
& \vspace{0.8mm} Median & \vspace{0.8mm} -0.50 & \vspace{0.8mm} 2 & \vspace{0.8mm} 34 & \vspace{0.8mm} 3.50 & \vspace{0.8mm} 19.50 \\[1mm]
\hline
\end{tabular}
\endgroup
\end{center}
% \vspace{2mm}
\vspace{-5.3mm}
\end{table}

Completing task \hyperlink{U1}{U1} required on average (median) 1.55 (1.50) seconds more with the tree editor than with the hybrid editor. 
% Correspondingly, completing task \hyperlink{U1}{U1} required a median duration that is higher by 1.50 seconds with the tree editor compared to the hybrid editor. 
Thus, for solving task \hyperlink{U1}{U1}, which requires locating model elements with specific values of interest, the hybrid editor had slightly better performance; however, no significant difference between the two exists. 
For task \hyperlink{U2}{U2}, the hybrid editor was faster on average by 315.95 seconds, indicating that using textual representations for expressing conditions improves comprehension significantly. 
In task \hyperlink{U3}{U3}, the tree editor was faster by 15.32 seconds than the hybrid editor; the median value outlines a higher difference, of 23 seconds. Concretely, the relative difference reported to the mean and median durations of the hybrid editor was 15\% and 27\%, respectively. This shows that tree editors are more appropriate to explore and analyse ordered lists of model elements, as navigating long ordered element lists in diagrams often incurs overheads. 
For task \hyperlink{U4}{U4}, the hybrid editor was faster by 10 seconds on average, however, this difference is negated by the delta median value, showing the existence of some outliers for this task. 
We conclude that no significant difference exists in performance between the editors when executing a task that requires finding references between elements. 
The hybrid editor yielded faster results in task \hyperlink{U5}{U5}, showing that finding elements that meet certain criteria is quicker in diagrams, as users have to meticulously expand and explore all elements in a tree editor.

Considering the modelling tasks, as indicated by the mean and median delta values of tasks \hyperlink{M1}{M1} and \hyperlink{M2}{M2}, we cannot conclude that there is a difference in performance between the editors for these tasks. 
For task \hyperlink{M3}{M3}, the hybrid editor resulted in improved performance, indicating that modifying existing conditions via a textual editor with syntax-aware editing features is quicker than through a tree editor. 
In task \hyperlink{M4}{M4}, the hybrid editor was faster on average by 8.94 seconds, however, the delta value of the median suggests a negligible difference between the performance of the editors. 
Thus, for solving a task that requires changing a reference, the editors have similar performance, although the hybrid editor may be slightly faster. For task \hyperlink{M5}{M5}, the hybrid editor was faster on average (median) by 9.51 (19.50) seconds, with a relative difference of 8\% (16\%), thus the hybrid editor was more effective in modelling conditions.

Finally, we used the non-parametric Wilcoxon signed rank test to perform statistical analysis between the duration taken by participants to complete the understanding and modelling tasks using the editors. Our statistical analysis with ($\alpha$=0.05) demonstrated significant differences in understanding tasks \hyperlink{U2}{U2} (p=0.00001) and \hyperlink{U5}{U5} (p=0.0127) and modelling tasks \hyperlink{M3}{M3} (p=0.0002) and \hyperlink{M5}{M5} (p=0.0329). These results are also underpinned by the boxplots in Figure~\ref{FIG-DURATION-BOXPLOTS}.

\vspace{1.2mm}\noindent
\textbf{Correctness.} 
All participants solved task \hyperlink{U1}{U1} correctly through the hybrid editor, whereas three participants solved the task incorrectly with the tree editor. From those, two participants reported an incorrect value as a solution, whereas one participant did not find a solution and abandoned the task. For task \hyperlink{U2}{U2}, two participants reported incorrect solutions with the hybrid editor, not being able to accurately state the left and right branches of the conditions represented textually. Correspondingly, seven participants solved the task incorrectly with the tree editor, as they constructed incorrect equivalent textual representations of the conditions from the tree editor by either grouping sub-conditions incorrectly, missing operators or using unsuitable operators. 
All participants solved correctly tasks \hyperlink{U3}{U3}--\hyperlink{U5}{U5} using both editors. 
The results evidence that the hybrid editor surpasses the tree editor in correctly locating specific model elements and in comprehending conditions.

In modelling tasks, one participant incorrectly performed task \hyperlink{M1}{M1} via the hybrid editor. The participant tried to set the name of a \metaclass{TestCampaign} by double-clicking its name on the diagram, and then entering the name in the pop-up menu displayed after double-clicking. The participant forgot a detail shown in the video tutorial: double-clicking specific elements in the hybrid editor instantiates new diagrams. As such, the participant instantiated a new diagram with the name entered in the pop-up menu, without noticing that the name of the \metaclass{TestCampaign} had not been set. Correspondingly, all participants solved task \hyperlink{M1}{M1} correctly through the tree editor. 

Task \hyperlink{M2}{M2} was solved correctly by all participants through both editors. One and six participants solved task \hyperlink{M3}{M3} incorrectly, with the hybrid editor and tree editor, respectively. The most encountered mistake made through the tree editor was that participants did not expand all condition elements to view their child elements, to be able to locate the operators that needed to be replaced. All participants solved task \hyperlink{M4}{M4} correctly with the hybrid editor, while only one participant failed using the tree editor. 
For task \hyperlink{M5}{M5}, two participants incompletely defined the required condition from scratch with the tree editor, while one participant made a mistake when textually defining the required condition; specifically, they used an \val{AND} operator instead of an \val{OR} operator. 
Consequently, we conclude that there is no significant difference between the editors for solving the tasks, with the exception of task \hyperlink{M3}{M3}, for which the hybrid editor yielded better results. %clearly superior. 
Fine-grained changes to conditions require a high level of attention to detail, and the results indicate that it is more appropriate to modify existing conditions through a textual representation rather than a tree representation.

\vspace{1.2mm}\noindent
\textbf{Errors Corrected. }
The errors corrected have been recorded for task \hyperlink{M5}{M5}, in which participants defined conditions from scratch. 
During the condition definition, 14 and 8 participants made incorrect actions with the hybrid editor and tree editor, respectively; the mistakes were corrected later in both cases.
% , which they later corrected. Correspondingly, 8 participants made mistakes when using the tree editor, that they later corrected. 
The results highlight that more than half of the participants faced challenges during modelling with the hybrid editor. 
The reason for the high number of mistakes is that the embedded textual editor from the hybrid editor allows any invalid input, such as typos. 
For all participants, in total, 8 errors have been corrected with the tree editor and 17 errors have been corrected with the hybrid editor. The mistakes made with the hybrid editor were minor, e.g, participants typed \val{LESS\_THAN} instead of \val{GREATER\_THAN}, typed lowercase \val{OR} instead of uppercase, and used curly braces instead of parentheses for defining \metaclass{BasicConditions}. However, the errors that have been corrected in the tree editor were of a higher magnitude. For example, participants created a \metaclass{BasicCondition} as a root element, then continued modelling the conditions, and later realised that the root should have been a \metaclass{CompositeCondition}. Next, participants deleted the \metaclass{BasicCondition}, and defined once again the conditions from scratch, as the tree editor does not allow replacing a model element.

\vspace{1mm}\noindent
\textbf{Clicks and Keystrokes.}
Table~\ref{tab:clicks_statistics} shows statistics for clicks and keystrokes, %. The number of clicks and keystrokes have been 
recorded for tasks \hyperlink{M3}{M3} and \hyperlink{M5}{M5}, which involve modifying and modelling conditions. The delta subtracts the mean and median number of clicks or keystrokes when using the hybrid editor from the corresponding values %mean and median number of clicks or keystrokes 
of the tree editor. 
On average, more mouse clicks (24.32) and fewer keystrokes (32.55) were used for solving task \hyperlink{M3}{M3} with the tree editor. Similarly, more clicks and fewer keystrokes were needed to solve task \hyperlink{M5}{M5} via the tree editor. 
The results are not surprising, %reflect what is typically expected, 
as hybrid editors require more typing to define textual expressions, which are defined via clicks through a tree editor. %However, the way in which the editors are used will dictate the results. For instance, there has been a participant who completed task \hyperlink{M3}{M3} using the tree editor with 21 clicks and 0 keystrokes, whereas there has been another participant who completed the same task via the hybrid editor with 21 clicks and 11 keystrokes, as many unnecessary clicks have performed.

\begin{table}[t]
\begin{center}
\begingroup
\renewcommand*{\arraystretch}{0.5}
\caption{\label{tab:clicks_statistics}Deltas of Mean and Median No. Clicks, Keystrokes}
\vspace{-3mm}
\begin{tabular}{|>{\centering}m{1.3cm}|>{\centering}m{1cm}|
>{\centering}m{1.65cm}|
>{\centering}m{1cm}|>{\centering\arraybackslash}m{1.65cm}|} 
\hline
\vspace{0.8mm}
$\mathbf{\Delta}$ $(\boldsymbol{T}-\boldsymbol{H})$ & \vspace{0.8mm} \textbf{{Clicks \hyperlink{M3}{M3}}} & \vspace{0.8mm} \textbf{{Keystrokes \hyperlink{M3}{M3}}} & \vspace{0.8mm} \textbf{{Clicks \hyperlink{M5}{M5}}} & \vspace{0.8mm} \textbf{{Keystrokes \hyperlink{M5}{M5}}} \\[3.25mm]
\hline
\vspace{0.8mm} Mean & \vspace{0.8mm} 24.32 & \vspace{0.8mm} -32.55 & \vspace{0.8mm} 29.70 & \vspace{0.8mm} -53.49 \\[1mm]
\hline
\vspace{0.8mm} Median & \vspace{0.8mm} 25 & \vspace{0.8mm} -14.50 & \vspace{0.8mm} 30 & \vspace{0.8mm} -48 \\[1mm]
\hline
\end{tabular}
\endgroup
\end{center}
% \vspace{2mm}
\vspace{-3mm}
\end{table}

\begin{table}[t]
\begin{center}
\begingroup
\renewcommand*{\arraystretch}{0.5}
%\vspace{-3mm}
\caption{\label{tab:confidence_statistics}Deltas of Mean and Median Confidence Levels}
\vspace{-3mm}
\begin{tabular}{|>{\centering}m{0.7cm}|>{\centering}m{1.3cm}|
>{\centering}m{0.77cm}|
>{\centering}m{0.77cm}|
>{\centering}m{0.77cm}|
>{\centering}m{0.77cm}|>{\centering\arraybackslash}m{0.77cm}|} 
\hline
\vspace{1.2mm} \textbf{{Type}} & \vspace{1.2mm} $\mathbf{\Delta}$ $(\boldsymbol{T}-\boldsymbol{H})$ & \vspace{1.2mm} \textbf{{\#1}} & \vspace{1.2mm} \textbf{{\#2}} & \vspace{1.2mm} \textbf{{\#3}} & \vspace{1.2mm} \textbf{{\#4}} & \vspace{1.2mm} \textbf{{\#5}} \\[1.8mm]
\hline
\multirow{2}{*}{\parbox{1\linewidth}{\vspace{2mm} \textbf{\hspace{2.3mm}U}}} & \vspace{0.8mm} Mean & \vspace{0.8mm} -0.68 & \vspace{0.8mm} -1.18 & \vspace{0.8mm} -0.14 & \vspace{0.8mm} 0.03 & \vspace{0.8mm} -0.05 \\[1mm]
\cline{2-7}
& \vspace{0.8mm} Median & \vspace{0.8mm} -0.50 & \vspace{0.8mm} -1 & \vspace{0.8mm} 0 & \vspace{0.8mm} 0 & \vspace{0.8mm} 0 \\[1mm]
\hline
\multirow{2}{*}{\parbox{1\linewidth}{\vspace{2mm} \textbf{\hspace{2.1mm}M}}} & \vspace{0.8mm} Mean & \vspace{0.8mm} -0.09 & \vspace{0.8mm} 0.09 & \vspace{0.8mm} -0.18 & \vspace{0.8mm} -0.34 & \vspace{0.8mm} -0.32 \\[1mm]
\cline{2-7}
& \vspace{0.8mm} Median & \vspace{0.8mm} 0 & \vspace{0.8mm} 0 & \vspace{0.8mm} 0 & \vspace{0.8mm} 0 & \vspace{0.8mm} 0 \\[1mm]
\hline
\end{tabular}
\endgroup
\end{center}
\vspace{-1mm}
\end{table}

\subsection{Confidence (RQ2)}
We answer research question \hyperlink{link_rq2}{RQ2} concerning the users' confidence when using both editors to do the specified tasks by analysing their self-reported confidence level, ranging between 1 (very low confidence) to 5 (very high confidence).
% As a reminder, research question RQ2 aims to evaluate the confidence of users when using both editors for carrying out various tasks. Recall that the confidence is a self-reported metric on a scale between 1 (very low confidence) to 5 (very high confidence).
%
Table~\ref{tab:confidence_statistics} shows statistics of the confidence level of participants about their task solutions.
As before, the table is split into tasks for understanding (U) and modelling (M), and the five numeric columns are linked to tasks \hyperlink{U1}{U1}--\hyperlink{U5}{U5} and \hyperlink{M1}{M1}--\hyperlink{M5}{M5}. The delta subtracts the mean and median confidence when using the hybrid editor from the mean and median confidence when using the tree editor. For task \hyperlink{U1}{U1}, participants were on average slightly more confident with their solution when using the hybrid editor, signifying  users' confidence
% This suggests that users are slightly more confident with their solutions when using a hybrid editor, 
for tasks that require understanding a domain and locating various model elements with specific values of interest. Participants were on average more confident by 1.18 points, with their solution to task \hyperlink{U2}{U2} when using the hybrid editor, signifying their confidence in 
% that users are more confident with their solutions when using a hybrid editor for 
understanding the meaning of conditions. For the remaining tasks, the results show no significant differences in the confidence of the solutions.

We also analyse the correlation between incorrect solutions and confidence level, aiming to determine how confident participants are with their incorrect solutions. A mean confidence of 3.13 and median confidence of 3 has been reported for the incorrect solutions to tasks \hyperlink{U1}{U1}--\hyperlink{U5}{U5} when using the tree editor. 
This shows that participants are neutral about the correctness of their partially incorrect solutions. Further, a mean and median confidence of 4.5 has been reported for the incorrect solutions to tasks \hyperlink{M1}{M1}--\hyperlink{M5}{M5} when using the tree editor. Similarly, a mean confidence of 4.67 and a median confidence of 5 has been reported for the incorrect solutions to tasks \hyperlink{M1}{M1}--\hyperlink{M5}{M5} when using the hybrid editor. These results show that participants were confident with their partially incorrect solutions, as they had not noticed any of their mistakes. This finding suggests that the confidence level is related to the ability of users to spot their mistakes. Participants could identify the understanding tasks they could not solve, and that was reflected in their confidence level for each understanding task. However, for the modelling tasks, the participants wrongly assumed that their solutions were correct, leading to a mistaken increased confidence level.

\subsection{Preference (RQ3)}
We answer research question \hyperlink{link_rq3}{RQ3} concerning the preferences of users with respect to the two editor types.
Before engaging in the study, participants were asked for their preferred editor type for modelling. As reported, 3 participants preferred a tree editor, 11 participants a hybrid editor, and 8 participants stated no preference. After performing the given tasks, the participants were asked again about their preferences. As reported, 2 participants preferred a tree editor, 16 participants (73\%) a hybrid editor, and 4 participants had no preference. The results show that one participant changed their preference from a tree editor to a hybrid editor, and half of the participants with no initial preference now prefer a hybrid editor.

Concerning understanding a testing space that does not contain condition expressions, 12 participants reported their preference for a diagrammatic editor, 7 participants for a tree editor, and 3 participants had no preference. 
To explore and navigate a testing space that does not contain condition expressions, 13 participants preferred a diagrammatic editor, 7 participants a tree editor, and 2 participants had no preference. 
To find relationships between elements of a testing space, 19 participants preferred a diagrammatic editor, 0 participants a tree editor, and 2 participants had no preference. 
To understand the meaning of condition expressions, 17 participants preferred a textual representation, 3 participants a tree representation, and 2 participants have no preference. 
For modelling conditions from scratch, 16 participants preferred a textual representation, 3 participants a tree representation, and 3 participants have no preference. 
To modify existing conditions, 17 participants preferred a textual representation, 4 participants a tree representation, and 1 participant has no preference. As reported, 20 participants stated that adding a textual syntax for modelling conditions inside the tree editor would improve usability and modelling efficiency; 2 participants did not share this opinion.

The feature of the tree editor that participants particularly liked was its conciseness and hierarchical representation that evidences the order and containment of elements. Participants also noted that the tree editor provided a better sense of control over the model, enabling faster exploration, especially when using the keyboard. However, several participants reported difficulties in understanding and defining complex conditions with the tree editor. Many participants would have liked the tree editor to use graphical symbols and colour-coding to differentiate between element types. 

Concerning the hybrid editor, participants particularly liked the flexibility in defining complex conditions and finding relationships between elements. Several participants liked the visual aspect of the hybrid editor, as they could relate colours and symbols to specific element types. 
However, some participants did not like the need to switch between multiple diagrams and considered that diagrams would not be readable for large models. Also, the textual syntax of conditions was not favoured by all participants, as it contained multiple kinds of brackets for grouping sub-conditions rather than a single kind, and it did not provide any formatting (pretty printing).

\subsection{Threats to Validity}
\label{SEC-THREATS}

% This study is subject to a number of threats to validity that are discussed in the following.

\vspace{0.8mm}\noindent
\textbf{Construct Validity. }
A construct validity threat is that participants may misinterpret task descriptions, although they could eventually understand the descriptions with the help of the session facilitator. 
If participants begin a task without completely understanding its description, they will need more time to solve it, and they will make more mistakes that they may later correct. 
Thus, the performance metrics for duration, correctness, and errors corrected are, by design, influenced by the clarity and understandability of task descriptions.

\vspace{1.2mm}\noindent
\textbf{Internal Validity. }
The main internal validity threat is that the study was conducted by accessing a remote computer via Team\-View\-er. 
For tasks \hyperlink{M3}{M3}--\hyperlink{M5}{M5}, participants used keyboard shortcuts, e.g., control+space to display an auto-completion menu in the textual editor from the hybrid editor. 
Some participants faced challenges, either because TeamViewer randomly did not trigger the actions associated with the keyboard shortcuts or because they did not know how to trigger keyboard shortcuts correctly due to their limited knowledge of the Windows OS. In such instances, durations reported with the hybrid editor were higher than expected.

\vspace{1.2mm}\noindent
\textbf{External Validity. }
An external validity threat is that the tree and hybrid editors have been evaluated with a single concrete syntax. 
Using a more appealing hybrid notation, with a different graphical representation and with a more intuitive textual syntax, may yield a better performance. Although the tasks covered a broad range of modelling activities, they were uncomplicated, as almost all participants could solve all tasks correctly. 
Further experiments would be useful to strengthen the generalisability of our findings, involving hybrid editors that are not Sirius/Xtext-based and other kinds of non-EMF or custom EMF-based tree editors.

\vspace{-2mm}
\section{Related Work}
\label{SEC-RELATED-WORK}
%\subsection{MDE for MRS and Testing}
% TODO: sort the papers into categories
% TODO: cut detail on long papers
\vspace{1.3mm}\noindent
\textbf{DSLs for Robotics.}
Model driven notations and toolchains have been used in 
RobotML~\cite{dhouib2012robotml}, BRICS~\cite{bruyninckx2013brics},
SmartSoft~\cite{schlegel2009robotic},
RoboChart~\cite{miyazawa2019robochart}, and in engineering robotic
perception and control
~\cite{ciccozzi2017engineering,schlegel2010design}. Developing
model-driven solutions for the robotics domain is an established area,
with a common mechanism for MDE engineering using finite state
machines and statecharts
\cite{gascuena2012model,paraschos2012model,skubch2011modelling}.
% Removing this to save space
%However, multi-robot interactions are under-studied, with
%\cite{cattivera2015model} surveying studies and finding only a
%minority dealing with MRS rather than single robots.
MRS interactions rather than single robots are studied
in \cite{cattivera2015model}. The FLYAQ family of graphical DSLs
~\cite{ciccozzi2016adopting} model the structure and behaviour of
multi-robot aerial systems, considering their spatial and
situational surroundings and tasks assigned to robots and their
behaviour. FLYAQ is extended with a specification language in~\cite{dragule2017generated}, enabling the definition of
domain-specific constraints in a declarative manner
and~\cite{pinciroli2016buzz} proposes a textual DSL for specifying the
behaviour of robot swarms.

\vspace{1.3mm}\noindent
\textbf{Testing DSL Tools.}
FuzzFactory ~\cite{padhye2019fuzzfactory} is a DSL for developing
domain-specific fuzzing applications. It generates code necessary for
fuzzing applications to communicate with a fuzzer, and obtain dynamic
domain-specific feedback during testing. 
VerifAI \cite{verifai-cav19} proposes a complementary
approach for tool-supported testing of cyber-physical systems,
typically integrated with AI and machine learning approaches.
%It uses a runtime monitor to assess the performance of the MRS, together with
%a search process to automatically find counterexamples.
VerifAI uses Scenic \cite{scenic-2019}, which presents a scenario
testing DSL, capable of specifying probabilistic features that can be
combined to generate test cases of interest.

\vspace{1.3mm}\noindent
\textbf{DSL Evaluation.}
FQAD~\cite{fqad-dsls-2013} introduces a methodology for DSL
evaluation, considering quality characteristics like expressiveness,
usability, and productivity to assess DSL suitability. 
%Using these characteristics and
%evaluator priorities, the process emits an assessment of DSL
%suitability. The technique was used in multiple case studies and
%found to be useful by stakeholders.
Despite its usefulness in multiple case studies, FQAD relies on
potentially subjective user perspectives, and also lacks
empirical validation in textual languages. Similarly, our empirical
study relies on subjective data such as user preferences; however, it
is also based on objective information, e.g., correctness, number of
clicks, errors corrected.

DSL evaluation for multi-agent systems is used in~\cite{mas-dsl-2016},
analysing language (elements and transformations), execution and
quality.  The results show that 80\% of the final product code could be
auto-generated.  However, it did not consider mistakes from language
users and their impact on learnability by new users.  The textual DSL
Athos~\cite{athos-2022} for specifying transport and traffic routing
simulations was evaluated to determine its effectiveness and
efficiency in model creation and understanding.  Athos was found
useful in modelling time and test scores when used by users
inexperienced in software development, and to a lesser degree, in more
experienced users.

User comprehensibility of a business use case between textual and
graphical flowchart (BPMN) representations is explored
in~\cite{ottensooser-compare-2010}. All user groups could
understand the textual representation (especially strong readers) and
experienced domain analysts obtained increased comprehension from
BPMN.
%Users with
%proxies for domain experience (business analysts) obtained increased
%comprehension from the graphical BPMN notation.
This suggests comprehension can be optimised by showing both
representations; first textual and then graphical. This may
not apply directly to our study where both representations are
significantly graphical, with the textual elements restricted to a
subset of the syntax (condition expressions).

In \cite{familiar-jaksic-2014}, tests were performed to determine
whether users could better create and manage feature models (FMs)
using a diagrammatic representation, as compared to a textual
representation. 
The results are aligned with our insights and showed statistically that users authored FMs of a higher quality and managed FMs with better efficiency and effectiveness with the diagrammatic representation.
Finally, in \cite{whispers-jolak-2020}, comparisons were performed between a graphical UML language and textual descriptions of the associated system, a familiar MVC fitness centre architecture. It found that the graphical representation is generally better than a textual description of the associated system, which broadly corroborates the findings in our investigation.
%

% \camera{In \cite{progeducation-saito-2017}, the effects of text-based and visual-based input on first-time programming learning were investigated, finding that the visual input method was more appropriate for an introduction to programming. However, the results are not comparable to those from our user study, due to different participant backgrounds (users in this study were unfamiliar with programming and formal grammar, whereas the participants from our user study were trained computer scientists and industry professionals).}

%\textbf{Hybrid DSLs}
%\lipsum[3]

% ACM's consolidated article template, introduced in 2017, provides a
% consistent \LaTeX\ style for use across ACM publications, and
% incorporates accessibility and metadata-extraction functionality
% necessary for future Digital Library endeavors. Numerous ACM and
% SIG-specific \LaTeX\ templates have been examined, and their unique
% features incorporated into this single new template.

% \vspace{-2mm}
\section{Conclusions and Future Work}
% \vspace{-1mm}
\label{SEC-CONCLUSIONS}
% We explored the tradeoffs between tree-based and hybrid graphical-textual model editors. 
% We executed an empirical user study in the context of an industrial use case, investigating the performance, confidence and preferences of users for both model editor types.  

Our empirical study, using an industrial use case, aimed to analyse and understand users' performance, confidence, and preferences between using tree-based and hybrid graphical-textual model editors.
% To understand the tradeoffs between tree-based and hybrid graphical-textual model editors, we conducted an empirical study using an industrial use case, analysing users' performance, confidence, and preferences for both model editor types.  
Concerning performance, there was a tradeoff between the speed of executing tasks using the editors.
% certain activities can be done faster via a hybrid editor, whereas other activities with a tree editor. 
% In particular, t
The hybrid editor was better for comprehending and modelling complex conditions, while the tree editor was better for analysing ordered lists of model elements. 
When modelling conditions, more corrected errors were found with the hybrid editor; however, the mistakes made were minor. 
Also, more keystrokes and fewer clicks were made for modelling conditions with the hybrid editor. 
Users were more confident with their solutions when using the hybrid editor to understand the meaning of conditions. 
Finally, before the study, 50\% of participants preferred a hybrid editor; at the end, the percentage increased to 73\%, showing an increased preference towards the hybrid editor.

We plan to validate the generality of our findings by using multiple graphical tree-based and hybrid syntaxes. Also, tree-based and hybrid editors should be evaluated with other domains and case studies, by carrying out a wider range of modelling activities.

\vspace{2.5mm}
% \begin{acks}
\noindent
\textbf{Acknowledgements.}
We thank our SESAME partners and the study participants. 
This work was supported by 
the SCHEME  InnovateUK project (\#10065634),
the  EU Horizon 2020 project SESAME (\#101017258), and
the Horizon Europe project SOPRANO (\#101120990).%. This work was also partially supported by 
%(Safety Critical Harsh Environment Micro-processing Evolution) 
% \end{acks}

\bibliographystyle{ACM-Reference-Format}
\bibliography{sample-base}

\end{document}